\documentclass[fleqn,10pt]{wlscirep}

\usepackage{breakurl}
\usepackage{adjustbox} 
\usepackage{amssymb} 
\usepackage{amsfonts}
\usepackage{amsmath}

\title{Beyond the network of plants volatile organic compounds}

\author[1]{Gianna Vivaldo}
\author[2*]{Elisa Masi}
\author[2]{Cosimo Taiti}
\author[3,4,5]{Guido Caldarelli}
\author[2]{Stefano Mancuso}

\affil[1]{National Research Council, Geosciences and Earth Resources (IGG), Pisa, Italy.} 
\affil[2]{Universit\`a di Firenze, Dipartimento di Scienze Produzioni Agroalimentari e dell'Ambiente (DISPAA) Viale delle Idee, 30 50019 Sesto Fiorentino Firenze.}
\affil[3]{London Institute for Mathematical Sciences, 35a South St. Mayfair W1K 2XF London UK.}
\affil[4]{Istituto dei Sistemi Complessi (ISC), Roma, Italy.}
\affil[5]{IMT School for Advanced Studies, Piazza San Francesco 19, 55100 Lucca, Italy.}
\affil[*]{elisa.masi@unifi.it}


\begin{abstract}
Plants emission of volatile organic compounds (VOCs) is involved in a wide class of ecological functions, as VOCs play a crucial role in plants interactions with biotic and abiotic factors. Accordingly, they vary widely across species and underpin differences in ecological strategy. In this paper, VOCs spontaneously emitted by 109 plant species (belonging to 56 different families) have been qualitatively and quantitatively analysed in order to classify plants species. By using bipartite networks methodology, based on recent advancements in Complex Network Theory, and through the application of complementary classical and advanced community detection algorithms, the possibility to classify species according to chemical classes such as terpenes and sulfur compounds is suggested. This indicates complex network analysis as an advantageous methodology to uncover plants relationships also related to the way they react to the environment where they evolve and adapt.
\end{abstract}

\begin{document}
\flushbottom
\maketitle
\thispagestyle{empty}
\noindent

\section*{Introduction} \label{sec:intro}
Plants produce an amazing variety of metabolites. Only a few of these are involved in “primary” metabolic pathways, thus common to all organisms; the rest, termed “secondary" metabolites, are characteristic of different plants groups\cite{theis2003evolution}. In fact, “secondary” metabolites, despite the name initially addressed to underline their inessiantiality for primary plant processes\cite{pichersky2000genetics}, are the result of different plants responses, through the course of evolution, to specific needs. Among such metabolites, volatile organic compounds (VOCs) play a dominant role\cite{dicke2010induced}. 
Being released by quite any kind of tissues\cite{dudareva2006plant, penuelas2001complexity} and type of vegetation (trees, shrubs, grass, etc.) as green leaf volatiles, nitrogen-containing compounds and aromatic compounds, plants VOCs can be emitted constitutively\cite{holopainen2010leaf,holopainen2010multiple} or in response to a variety of stimuli. They are in fact involved in a wide class of ecological functions, as a consequence of the interactions of plants with biotic and abiotic factors\cite{spinelli2011emission}. Plants use VOCs to perform indirect plant defence against insects\cite{mumm2003chemical}, to attract pollinators\cite{dudareva2000biochemical}, for plant-to-plant communication \cite{baldwin2006volatile, heil2010explaining}, for thermo-tolerance and environmental stress adaptation (see more references in\cite{holopainen2010multiple}), to defend from predator\cite{war2012mechanisms}. 

According to their biosynthetic origin and chemical structure, plant volatiles can be grouped into isoprenoids or terpenoids, but also oxygenated VOCs (OVOCs), such as methanol (CH$_{4}$O), acetone (C$_3$H$_6$O), acetaldehyde (C$_2$H$_4$O), methyl-ethyl-ketone (MEK, C$_4$H$_8$O) and methyl-vinyl-ketone (MVK, C$_4$H$_6$O)\cite{ruuskanen2009measurements}; in few cases, sulfur compounds (e.g. in Brassicales) and furanocoumarins and their derivatives (e.g in Apiales, Asterales, Fabales, Rosales) are found\cite{agrawal2011current,berenbaum2008facing}.

Interestingly, VOCs emissions strongly depend on the species (see \cite{llusia2002seasonal} for references). Indeed, different plant lineages often adopt different chemical solutions to face the same problem; this is the case, for example, of the different odorous volatiles emitted by different flowers for solving the common problem of attracting the same type of pollinator, which usually visit a large amount of plant species\cite{pichersky2000genetics}.

In this paper we apply both complex networks analysis\cite{caldarelli2007scale,Boccaletti2006175,barrat2004architecture} and community detection\cite{raghavan2007near,newman2004finding} to identify an eventual hierarchy among the available species, on the basis of their similarities in terms of VOCs emissions. Complex Network Theory\cite{ma2016wiener,li2013note,cao2014extremality} has been already successfully used in ecology to determine, for example, the stability and robustness of food webs\cite{Dunne01102002} with respect to the removal of one or more individuals from the network, or in biology to study the structure of protein interactions in the cell by the so-called protein interaction networks (PINs)\cite{Stelzl2005957}. Moreover metabolic networks are used to study the biochemical reactions which take place into living cells\cite{proulx2005network}. Still, biological networks found important applications in medicine\cite{barabasi2011network}, where they are applied as a solution to human diseases comorbidity analysis\cite{leecomorbidity}, or to study the structural and functional aspects of human brain, by defining the reciprocal interactions of the cerebral areas\cite{stephan2000computational}. Nevertheless, the application of Complex Networks Theory in botany is still scarce, exception made for some tentatives of comparing different ecosystems looking for steady (i.e. ``universal'') behaviours\cite{caretta2008}. Recent applications of graph theory in botany deal with the attempt of assessing plants species similarities on the basis of both their diaspore morphological properties, and fruit-typology ecological traits\cite{vivaldo2016networks}. Following the same approach, in this paper we perform network analysis with the goal of identifying communities of ``similar'' species, starting from the volatiles they emit, or more generally from the ways the species share between them to react to external wounding stimuli. 

At this purpose, data are represented by means of bipartite graphs, which are particularly suitable to study the relations between two different classes of objects and to group individuals according to the properties they share. More in details, the vertices of a bipartite graph can be subdivided into two disjoint sets, such that every vertex of one set is connected only with a vertex of the other set. No links are present between vertices belonging to the same set. In our case, the plants species and the volatiles they emit will define the two independent sets of vertices of the bipartite graph built from botanical data. In practice two different graphs can be analysed: the first one is made up by all plants species (as vertices) connected on the basis of the common properties they share, i.e. in that case the amount of VOCs emitted; the second one is made up by VOCs connected accordingly to the plants that share the same emissions. Once the graph are suitably deduced from the experimental data, community detection is a powerful method to classify in a quantitative way the different species creating a taxonomic tree\cite{newman2004finding}.

The paper is organized as it follows: in Section ``Results'' we present the main outputs of the analysis conducted on the dataset considered; in Section ``Discussion'' we debate the main implications of our results and we propose possible further developments of the actual study. We refer the reader interested in more details about both data and methodology to Section ``Materials and Methods''.


\section*{Results and Discussion} \label{sec:results}
The present research work focuses on a group of $75$ volatiles emitted by $109$ different plant species in basal conditions, in order to understand if taxonomy-related plants emit a similar VOC composition. To assure the analysis to be robust and consistent, we measured the volatiles emitted by each plant species by a three times replication experiment. We refer to the section ``Materials and methods'' for a suitable description of the dataset preparation.

Complex networks analysis is applied to the VOCs dataset represented as bipartite network, in order to easily define metrics and hidden statistical properties able to discriminate and classify plant taxonomy based on VOCs patterns. 

\subsection*{Data preprocessing}
The $109$ plants species analysed are representative of $56$ families, and the dataset is quite homogeneous in terms of families percentages. The most copious families are: \textit{Asteraceae} ($8.26\%$), \textit{Solanaceae} ($6.42 \%$), \textit{Rosaceae} ($6.42 \%$), \textit{Fabaceae} ($5.5 \%$), \textit{Brassicaceae} ($4.59 \%$), and \textit{Polygonaceae} ($3.67 \%$). All the other families are present at lower percentages.   

\begin{figure}
\centering
\includegraphics[width=1\textwidth]{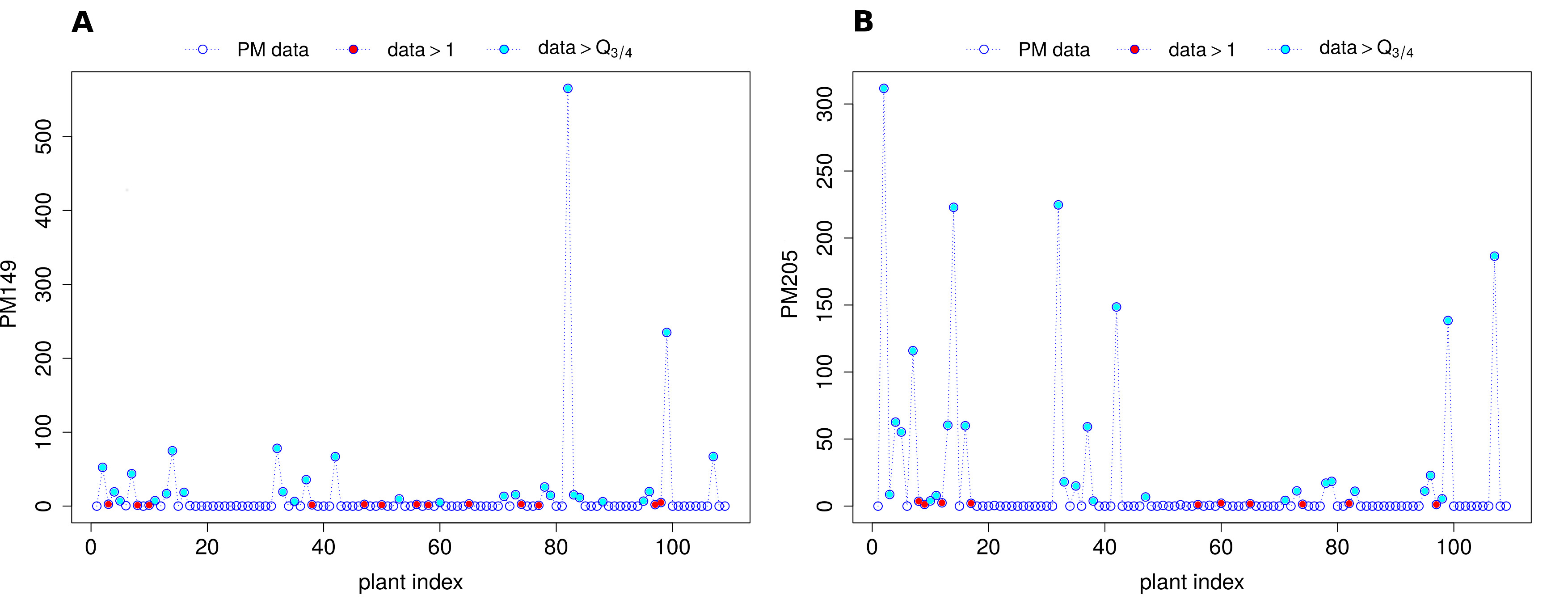} 
\caption{PM$149$ (Tp/STp-f) (panel A) and PM$205$ (STp) (panel B) emissions. Protonated mass data are represented by empty blue bullets. Red dots correspond to values larger than 1, while cyan dots refer to those data exceeding $Q_{\frac{3}{4}}$. x-axis: plants index. Protonated masses are expressed as mass-to-charge (m/z) ratios.} 
\label{fig:data}
\end{figure}

To evaluate the data statistical structure we plotted for each protonated mass the emission recorded for all the $109$ plants species. Figure~\ref{fig:data} (empty blue bullets) shows the emission of protonated masses PM$149$ (panel A) and PM$205$ (panel B), as two examples of VOCs records behaviour. Protonated masses are expressed as mass-to-charge (m/z) ratios. From the chemical composition point of view, PM$149$ and PM$205$ belong to terpenes/sesquiterpenes fragments (Tp/STp-f) and sesquiterpenes (STp) classes, respectively. The VOCs series turn out to be characterized by the superposition of an irregular, abruptly changing pulsatile component and a slowly changing one. More in details, zero-values indicate the lack of emission of that specific VOC for the corresponding plants, and the flat and uniform plateau suggests a small emission of the same VOC. Finally, spike-like pulses, clearly emerging from the background, are related to a huge emission of that VOC for a given plant. Figure~\ref{fig:data} suggests that both the protonated masses PM$149$ and PM$205$ are emitted in large quantity just by few species. That behaviour turned out to be representative of the whole dataset (not shown). 

\begin{figure}
\centering
\includegraphics[width=0.8\textwidth]{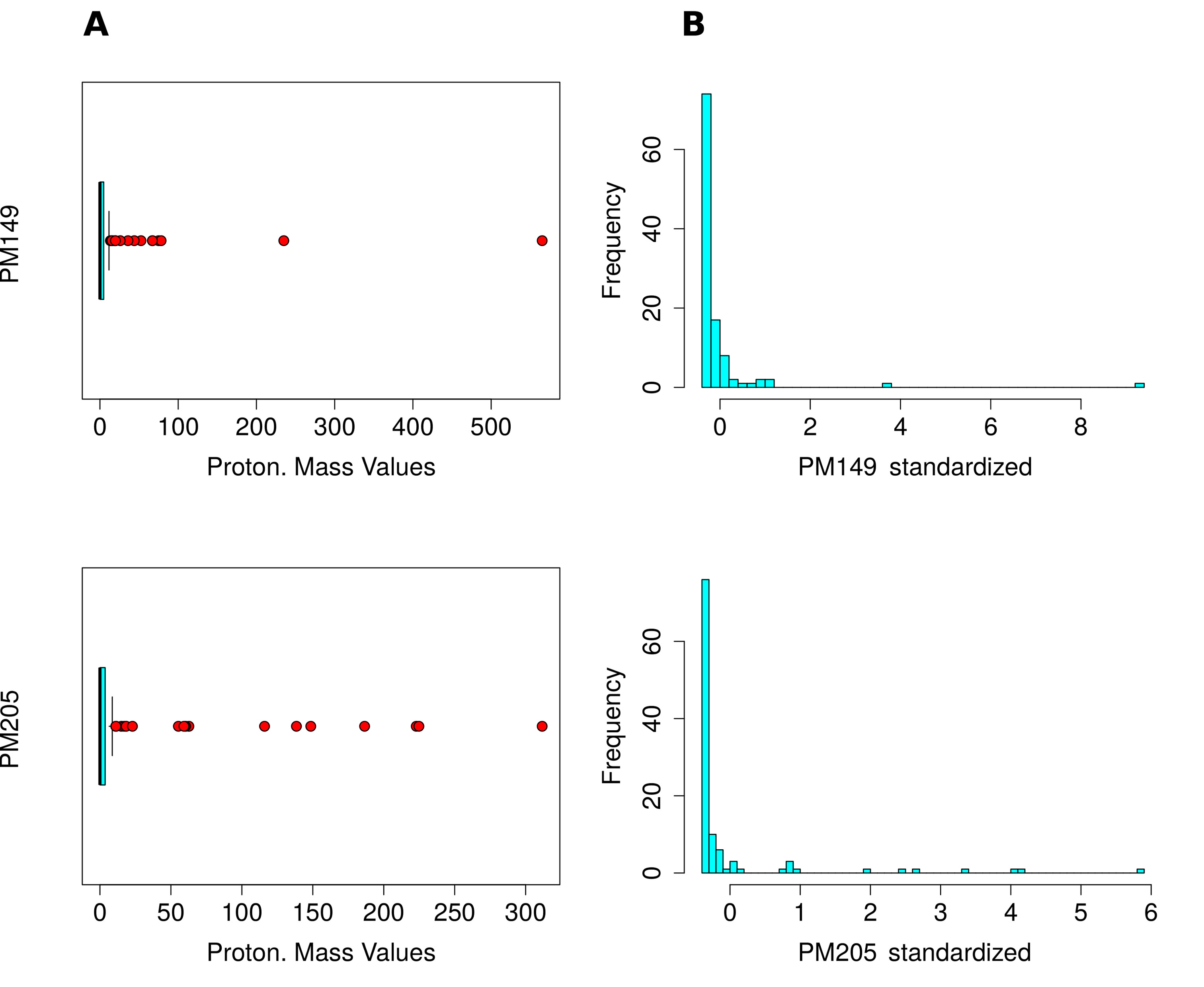}
\caption{PM$149$ (Tp/STp-f) and PM$205$ (STp) emissions. Panels A: boxplots, $IQR$ (cyan rectangles) and outliers (red dots). Panels B: Absolute frequency histograms (y-axes) versus data standardized values (x-axes), obtained by removing the sample mean, and by normalizing the residuals to the sample standard deviation. Protonated masses are expressed as mass-to-charge (m/z) ratios.} 
\label{fig:boxplot}
\end{figure}

From a statistical point of view the same result was confirmed by the presence of outliers inside each record, that can be easily visualized by boxplot methodology\cite{tukey1977exploratory,vandervieren2004adjusted} (see Section ``Materials and methods'' for more details). Outliers are shown in Fig.~\ref{fig:boxplot} (panels A, red dots), and they correspond to those observations far from the sample mean. In that case, since the behaviour was coherent for all the VOCs, we excluded the presence of outliers as a consequence of merely experimental errors. Rather, protonated mass records were characterized by heavy-tailed distribution, as Fig.~\ref{fig:boxplot} (panels B) shows: few values lie in the queues of the absolute frequencies sample data histograms. Standardized values were employed in order to assure the results comparability. 

Notwithstanding the clear dominating behaviour of some species emissions with respect to the other plants, for a given VOC, the statistical procedure of taking into account just the highest recorded values (extreme values) turned out to be too restrictive. In fact, not even a small emission of a protonated mass can be neglected from an experimental point of view. A low emission is as well a signal from a wounded leaf, and it has to be taken carefully into account when comparing the several species reciprocal behaviour with respect to an external wounding perturbation.

\subsection*{Basic network analysis}
We considered two different ways of building the plants network, depending on the statistical measure used to represent the highly not-gaussian behaviour of the series. In the first case, we set a fixed threshold for the signal intensity ($1$  normalized counts per second, \textit{n}cps) and we considered significant all the emissions larger than it (graph: $G_1(V,E)$). In the second case, we applied a more severe criterion, and we decided to take into account just the emissions above the third quartile of the corresponding data statistical distribution, i.e. $Q_{\frac{3}{4}}$ (graph: $G_2(V,E)$). Figure~\ref{fig:data} shows both the approaches applied to PM$149.1$ (panel A) and PM$205.1$ (panel B). Red dots in both panels highlight values larger than $1$, while cyan bullets represent the value exceeding $Q_{\frac{3}{4}}$.

In both cases, a bipartite network was build, made up by $V = 184$ vertices, subdivided into two layers: the first one made up by $V_P = 109$ plants species and the second one composed by $V_{PM}=75$ emitted VOCs. By definition of bipartite graph, connections were possible between vertices belonging to the two different layers, only. No links are present among plants, as well as among VOCs. Plants species networks are subsequently defined by considering as vertices the plant species in the database, so as bipartite projections of both $G_1(V,E)$ and $G_2(V,E)$. Two vertices are connected if they share at least one common property, in other words, if they emit almost the same amount of a specific VOC. 
For every network, we considered size (number of edges), order (number of vertices), degree (average and its distribution), density (the ratio of actual vertices against the possible ones), clustering and finally the community structure. 

\subsubsection*{Threshold-based graph}
The plants graph corresponding to the first method was created as a bipartite projection of graph $G_1(V,E)$. In the resulting graph $G^P_1(V_1,E_1)$ plants are interconnected on the basis of the common VOCs they emit. $G^P_1(V_1,E_1)$ is made up of $V_1=V_P=109$ vertices (plant species), and $E_1 = 5,886$ edges. Species $i$ and species $j$ are linked if they share at least one common emitted protonated mass. The weight $w_{ij}$ of each link $e_{ij}$ is given by the total number of shared VOCs between species $i$ and species $j$. $G^P_1(V_1,E_1)$ is a fully connected graph, its density $D = \frac{2E_1}{V_1(V_1-1)}$ is equal to 1, and the degree of each node is equal to $108$, which is also equal to the nodes mean degree ($\overline{k}=\frac{1}{V_1}\sum_{i=1}^{V_1}k_i=\frac {2E_1}{V_1}$). Each vertex is connected to all the other vertices, or equivalently each species emits at least one VOCs in common with all the other species. 

That network structure is poorly able to extract information about the dominant behaviour of one species with respect to the others, in terms of their emissions. 
Concerning the links weights distribution, the maximum number of protonated masses shared by two species is $66$, and in average species are connected by links of weight $w_{ij}=24$, in agreement with the dense structure of the network.

\subsubsection*{Third-quartile-based graph}
The plants graph corresponding to the second test was analogously constructed as the species-vs-species bipartite projection graph of $G_2(V,E)$ graph. Again, the common emitted VOCs determine the presence or not of a (weighted) link between two nodes. $G^P_2(V_2,E_2)$ is made up by $V_2= V_P=109$ vertices and $E_2 = 2,343$ edges. Links are less by construction: in that case, for each VOC just the emissions larger than $Q_{\frac{3}{4}}$ were considered significant. It follows that the network construction procedure accounted for a more severe pruning. Graph density reduces to 0.39 consistent with the fact that the graph is not fully connected. Rather, isolated vertices emerge, suggesting the presence of plants which do not emit any of the measured VOCs at a high level. By removing them the graph density increase to $0.73$.

\begin{figure}
\centering
\includegraphics[width=1\textwidth]{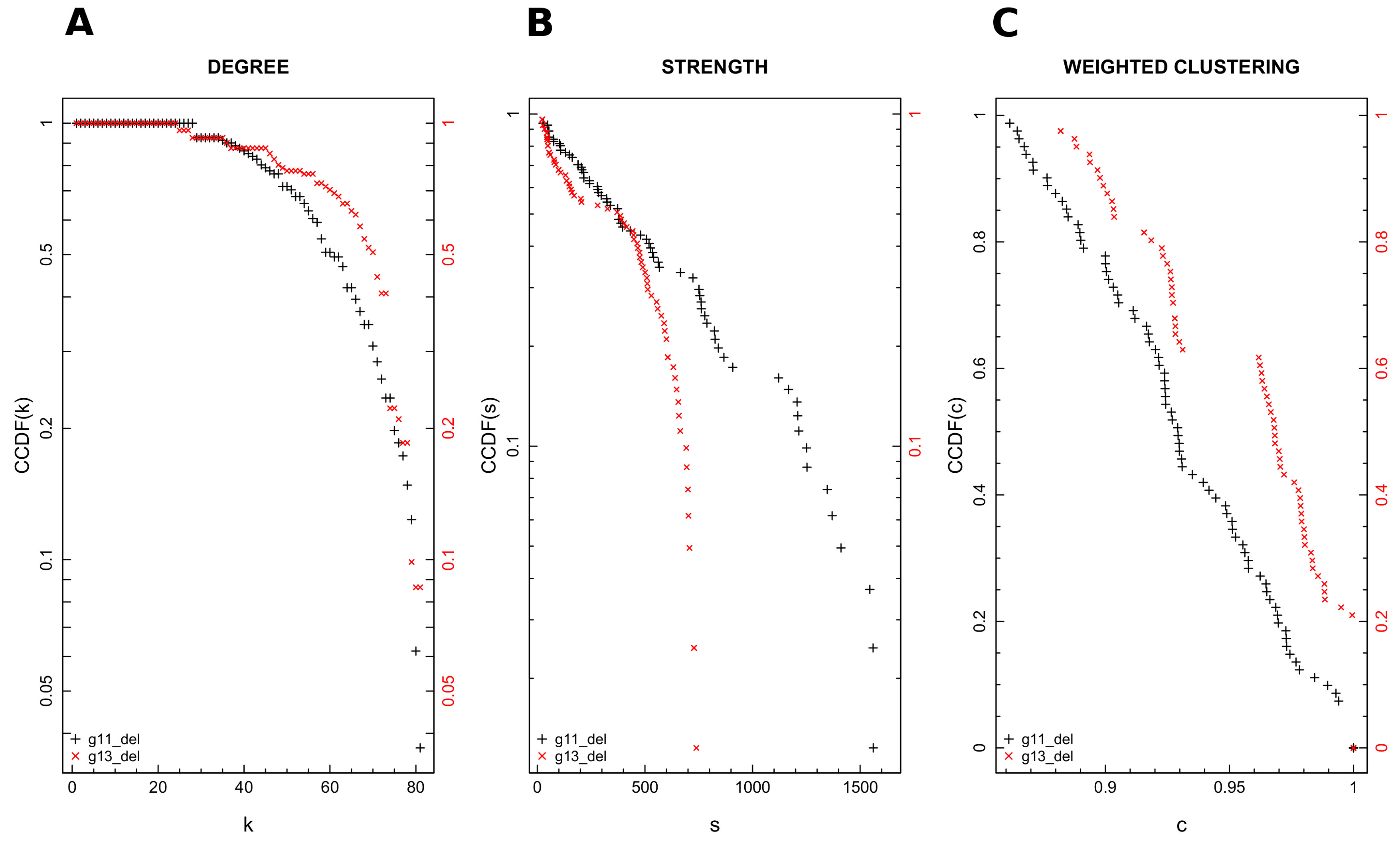}
\caption{Complementary cumulative distribution functions (CCDF) of degree and strength are reported in log-line scale in panel A and B, respectively, for both $G^P_2(V_2,E_2)$ (black crosses) and $G^P_3(V_3,E_3)$ (red crosses). Panel C, moreover, shows weighted clustering coefficient distribution. More precisely, CCDF (on y-axis) is plotted versus the weighted clustering parameter (x-axis) on linear scale. The isolated nodes were not taken into account for the corresponding network's basic metric analysis.} 
\label{fig:network_metrics}
\end{figure}

In that case the majority of species share few common VOCs emitted (i.e. the mean of the edges weights is around $5$). On the contrary some vertices are connected by heavy links (the maximum weight's value is $67$, similarly to the previous case).  Figure~\ref{fig:network_metrics} (panel A, black crosses) shows the network degree distribution $P(k)$, representing the fraction of vertices with degree $K > k$. A log-line plot is chosen to display the degree complementary cumulative distribution function (CCDF). The graph strength distribution is also shown in Fig.~\ref{fig:network_metrics} (panel B, black crosses) in log-line scale. The strength $s$ of a vertex corresponds practically to its weighted degree, thus it takes into account the total weight of the vertex connections, and it allows one to identify high and low concentration edges-regions inside an undirected graph. The maximum strength value is equal to $s_{max}=1,624$ and it corresponds to \textit{Lavandula spica} L. (Lavender) species, the minimum is equal to $s_{min}=27$ and it is common to \textit{Humulus lupulus} L. (Wild hop), \textit{Actinidia arguta} (Siebold \& Zucc.) (Hardy kiwi), \textit{Ficus benjamina} L. (Weeping fig), \textit{Magnolia liliiflora} (Desr.) (Japanese magnolia), and \textit{Diospyros lotus} L. (Date-plum) species. 
Finally, Fig.~\ref{fig:network_metrics} (panel C) shows the local clustering coefficient, defined as the tendency among two vertices to be connected if they share a mutual neighbour. 

Taken as a whole, Fig.~\ref{fig:network_metrics} suggests that plants network is not dominated by some central nodes with a huge amount of connections linking them to all the other minor vertices. Notwithstanding, some species emit a large quantity of VOCs and communities detection algorithms are applied to identify them and the respective aggregating VOCs. The graph $G^P_2(V_2,E_2)$ isolated nodes were removed before performing that basic metrics analysis for visual reasons. The degree and strength of an isolated node are equal to $0$ by definition and the clustering coefficient is not defined.

\subsubsection*{Selected-VOCs graph}
A third test was performed on a reduced version of the original database. Certain VOCs which could be more strictly associated to the mechanical wounding performed during the sample measurements than to plant species-specific emissions were excluded. Indeed, certain compounds such as methanol, acethaldeyde, some C6-compounds, etc. \cite{loreto2006induction,brilli2011detection} are produced by almost all plant species, but there is no a common behaviour in terms of quality and quantity of VOCs involved \cite{degen2004high,wu2008comparison}; their inclusion in the database could lead to misinterpretation. Furthermore, other compounds that turned out to be less powerful in the aggregation features, as highlighted by the above described analyses, were removed from the dataset. 
As a result, a selection of $30$ protonated masses were taken into account.

In order to compensate that filter introduced by the hand-made choice of the relevant VOCs to be considered for the analysis, a threshold equal to $0$ was used to distinguish between relevant and negligible emissions of that specific VOC. 
The corresponding bipartite network $G_3(V,E)$ was made up by $V = 139$ vertices subdivided in two sets: $V_{P} = 109$, analogously to previous graphs, and $V_{PM} = 30$. In order to study plants network, the bipartite projection $G^{P}_3(V_3, E_3)$ was analysed. The vertices are still $V_3=V_P=109$, while the edges are equal to $E_3=2,522$, similarly to the third-quartile-based graph. The graph density is $0.43$ due to the presence of $28$ isolated nodes, while it raises to $0.78$ if they are removed. 
Concerning the graph basic metrics, Fig.~\ref{fig:network_metrics} (panel A, red crosses) shows $G^P_3(V_3,E_3)$ complementary cumulative degree distribution $P(k)$, while Fig.~\ref{fig:network_metrics} (panel C, red crosses) depicts the graph strength distribution. Both figures are in log-line scale. 
The network strength maximum value decreases to $s_{max}=746$, but it still corresponds to \textit{Lavandula spica} L. (Lavender) species, which again emerges as the most connected node. On the other side, the strength minimum value is $s_{min}=23$ for \textit{Cyperus papyrus} L. (Papyrus), \textit{Salicornia europaea} L. (Glasswort), and \textit{Solanum quitoense} Lam. (Naranjilla) species. Further, $G^P_3(V_3,E_3)$ is characterized by a smaller range of strength values with respect to $G^P_2(V_2, E_2)$, and a more restricted set of nodes seem to dominate the network behaviour. Nevertheless, the graph degree and strength distribution do not suggest the presence of a scale-free structure behind our data. Finally, Fig.~\ref{fig:network_metrics} (panel C, red crosses) shows $G^P_3(V_3,E_3)$ clustering coefficient. The behaviour is similar to the one observed for $G^P_2(V_2, E_2)$ graph. Such as for $G^P_2(V_2,E_2)$ graph, isolated nodes were removed before performing that basic metrics analysis. Analogously, the strength minimum value is performed after excluding the isolate nodes, since the degree $k$ and thus the strength $s$ of an isolated node are equal to $0$ by definition.

\subsection*{Community detection analysis} 
\subsubsection*{Threshold-based and third-quartile-based graphs}
A first attempt to group plants on the basis of the VOCs emitted was performed by applying the community detection to both the dense $G^P_1(V_1,E_1)$ graph and the third-quartile-based graph $G^P_2(V_2,E_2)$. For both of them, subgraphs were obtained filtering-out a growing number of links, from the lower to the higher weighted ones. A unit-based normalization was applied to edges weights to limit their values to the [0, 1] range ($w^{resc}_{ij}$ parameter in Tab~\ref{tab:g9_clus}). Four communities detection algorithms were applied: (i) Louvain or Blondel’s modularity optimization algorithm (BL), (ii) fast greedy hierarchical agglomeration algorithm (FG), (iii) walktrap community finding algorithm (WT), and (vi) label propagation community detection method (LP). We refer to the section ``Materials and methods'' for a detailed description of the communities detection methods. 
\begin{table}
\centering
\begin{tabular}{|c|c|c|c|c|c|c|c|c|c}
\hline
\hline
FG & WT & BL & LP & $w^{resc}_{ij}$ & E    & N   & is.connected & density \\ \hline
2  & 2  & 2  & 1  & 0               & 5886 & 109 & TRUE         & 1       \\
2  & 2  & 2  & 1  & 0.1             & 5303 & 104 & TRUE         & 0.99    \\
2  & 2  & 2  & 1  & 0.2             & 4776 & 101 & TRUE         & 0.95    \\
2  & 2  & 2  & 1  & 0.3             & 3220 & 85  & TRUE         & 0.9     \\
2  & 2  & 2  & 1  & 0.4             & 1316 & 58  & TRUE         & 0.8     \\
2  & 1  & 3  & 1  & 0.5             & 697  & 44  & TRUE         & 0.74    \\
2  & 28 & 2  & 1  & 0.6             & 309  & 28  & TRUE         & 0.81    \\
2  & 4  & 2  & 1  & 0.7             & 156  & 21  & TRUE         & 0.74    \\
2  & 12 & 2  & 1  & 0.8             & 48   & 12  & TRUE         & 0.73    \\
2  & 3  & 2  & 1  & 0.9             & 13   & 7   & TRUE         & 0.62    \\
1  & 2  & 1  & 1  & 1               & 1    & 2   & TRUE         & 1       \\
\hline 
\hline	
\end{tabular}
\caption{Communities detection of graph $G^P_1(V_1,E_1)$ by fast greedy (FG), walktrap algorithms (WT), Blondel modularity optimization (BL), and label propagation (LB). Several filtered-by-edges-weight versions of the graph were analysed (one for each row). Graph edges weight values are normalized to the interval $\left[0,1\right]$.}
\label{tab:g9_clus}
\end{table}

Notwithstanding some discrepancies in the results depending on algorithms optimization after pruning the network, two big communities emerge from $G^P_1(V_1,E_1)$ analysis, which turned out to be robust to algorithm changes and to the filtering procedure of the edges weights (see Tab.~\ref{tab:g9_clus}), exception made for severe filters (rescaled weight parameter $w_{ij} > 0.5$ in Tab.~\ref{tab:g9_clus}). In that case, almost half of the graph nodes were filtered out, thus reducing the reliability of the related results as the consequence of a huge loss of information. On the contrary, by pruning the graph from the most heavy links, the results were statistically comparable thus meaning that the plants network was not dominated by some big vertices acting as hubs of the whole system. The two uncover communities embed the $61.47\%$ and $38.53\%$ of the total amount of species inside the database, respectively.

\begin{figure}
\centering
\includegraphics[width=1\textwidth]{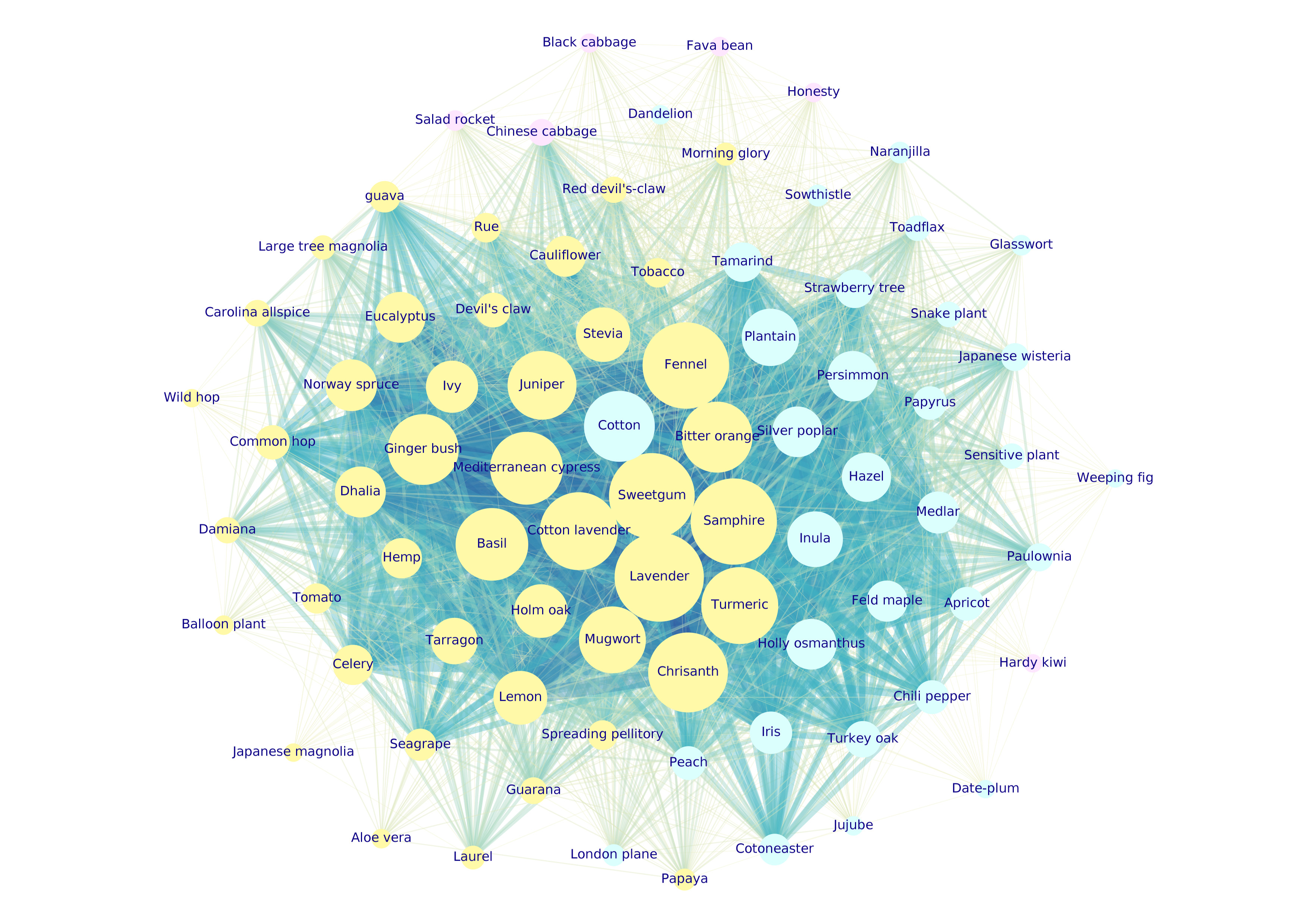}
\caption{$G^P_2(V_2,E_2)$ third-quartile-based graph. Each color corresponds to one detected community: cluster $1$ (yellow), cluster $2$ (aqua), cluster $3$ (violet). The $28$ isolated nodes are not shown (cluster $4$). Nodes dimension is proportional to nodes weighted degree. Edges thickness is proportional to the edges weight. \textit{Lavandula spica} L. (Lavander), \textit{Foeniculum vulgare} Mill. (Fennel),  \textit{Crithmum maritimum} L. (Samphire), \textit{Liquidambar styraciflua} L. (Sweetgum), visible as biggest yellow nodes, are some of the most active species in terms of VOCs emissions.} 
\label{fig:g11delQ3}
\end{figure}

The situation improved by analyzing the communities of $G^P_2(V_2,E_2)$ graph. Figure~\ref{fig:g11delQ3} is a representation of $G^P_2(V_2,E_2)$ plants network. The dimension of each node is proportional to the node's weighted degree. The thickness of each link connecting two nodes $i$ and $j$ is proportional to the link's weight, $w_{ij}$. Nodes colours refer to cluster membership.
In that case two big clusters emerge from a basic community detection. They embed $44$ and $31$ species, i.e. respectively the $40.4\%$ and $28.4\%$ of the species present in the dataset (yellow and aqua clusters in Fig.~\ref{fig:g11delQ3}). \textit{Brassicaceae} family started to be pretty grouped in a third small family ($6$ species only accounting for the $5.5\%$ of the species dataset, violet cluster of Fig.~\ref{fig:g11delQ3}), exception made for the \textit{Brassica oleracea} L. var botrytis species (Cauliflower) which belongs to another community (yellow cluster in Fig.\ref{fig:g11delQ3}).
By construction $28$ isolated nodes emerged (not shown in Fig.~\ref{fig:g11delQ3}), corresponding to species which were not sharing any of the measured VOCs with the other plants. Isolated nodes accounted for the $25.7\%$ of species total amount. Again, the results were consequent to the simultaneous application of more than a single methodology. The findings proved to be independent from the applied methodology and they were considered robust and reliable from a statistical point of view. 
Hereafter, the composition of every cluster is summarised, together with the protonated mass that the species share at graph's communities level:
\begin{itemize}
\item{cluster $1$}: $31$ species ($28.4\%$ of the database total species) grouped in $21$ families; prevailing families: \textit{Rosaceae}, \textit{Asteraceae}, \textit{Fabaceae}, \textit{Ebenaceae}, \textit{Plantaginaceae}, and \textit{Solanaceae}. Two VOCs in particular are responsible for that partitioning: PM$27$ (hydrocarbons, Hyd) and PM$73$ (acids, A) ($20$ species), followed by PM$55$ (aldehydes fragment, Ald-f), PM$89$ (esters, E), PM$115$ (acids, A) ($19$ species), and PM$53$ (fragment, f), PM$81$ (aldehydes fragments, Ald-f) ($18$ species). In general, the more informative VOCs for this cluster are compounds belonging to several chemical classes. Notice that from $m/z = 123$ (PM$123$) to $m/z = 205$ (PM$205$), where peaks deriving from terpenes, sesquiterpenes and their fragments are found, the emissions are null for all the species. One species can emit more than one VOC, so that all the species can be counted more than once to assess how many species share the same protonated mass emission. 
\textit{Gossypium herbaceum} L. (Cotton), \textit{Plantago lanceolata} L. (Plantain), and \textit{Inula viscosa} L. (Inula) species are between the highest weighted degree nodes in Fig.~\ref{fig:g11delQ3}.

\item{cluster $2$}: it is the biggest community, made up of $44$ species ($40.4\%$ of the total species amount) grouped into $27$ families; dominant families: \textit{Asteraceae}, \textit{Apiaceae}, \textit{Cannabaceae}, \textit{Lamiaceae}. The species belonging to that cluster emit, taken as a whole, a large amount of VOCs. They share in particular the emission of VOCs which are or refer to terpenes compounds, which are among the principal odour-like molecules emitted by plants flowers and leaves. In details, 28 species share PM$123$ and PM$135$, both terpenes or sesquiterpenes fragments (Tp/STp-f); 27 species share PM$93$ (Tp-f), PM$95$ (STp-f), PM$105$ (heterocyclic aromatic compounds, HeArC), PM$109$ (Tp-f), PM$119$ (Tp-f), PM$121$ (Tp-f), PM$137$ (Tp/STp-f), PM$143$ (ketones and aldehydes, K/Ald), PM$149$ (Tp/STp-f), PM$163$ (STp-f), PM$205$ (STp); $26$ species share PM$91$ (hydrocarbons, Hyd), PM$107$ (HeArC), PM$111$ (aldehydes, Ald), PM$153$ (Tp-f). 

Accordingly, that community includes plant species characterized by intense flavour, such as \textit{Lavandula spica} L. (Lavander, a well known plant used for its flavour), \textit{Foeniculum vulgare} Mill. (Fennel, an anise-flavored spice), \textit{Crithmum maritimum} L. (Samphire, a very flavoured sea fennel), and \textit{Liquidambar styraciflua} L. (Sweetgum, commonly used as flavor and fragrance agent). A more detailed description of cluster $2$ is supplied hereafter.

\item{cluster $3$}: $6$ species only ($5.5\%$ of total species) from $3$ families: \textit{Brassicaceae} (dominating family with $4$ species), \textit{Actinidiaceae}, and \textit{Fabaceae}. Interestingly, the \textit{Brassicaceae} Cauliflower belongs to the previous community (i.e., to cluster $2$, where species characterized by more intense odours and presence of terpenes compounds are clustered). Indeed, Cauliflower is, among the \textit{Brassicacaeae} species included in the present study, one of the richest in VOCs and terpenes \cite{van1991identification, geervliet1997comparative}. This is the most homogeneous community in terms of family composition. PM$63$, a typical sulfur compound (SC), is the most emitted VOC, being released by $5$ species ($4$ of them belonging to the \textit{Brassicaceae} family), followed by another sulfur compound, PM$49$, and PM$83$ (alcohols fragment, Alc-f) ($3$ species), PM$87$ (Ald/Alc). In particular \textit{Brassica rapa} L. (Chinese cabbage) emits also PM$85$ (Alc-f), PM$103$ (esters, E), PM$117$ (Alc), PM$129$ (Alc), PM$143$ (ketones and aldehydes, K/Ald). 
The latter protonated mass, tentatively identified as 2-Nonanone \cite{buhr2002analysis} has been already reported in Chinese cabbage\cite{pierre2011differences}.
The emission of all the other VOCs is null for the whole species set. 

\item{cluster $4$}: $28$ isolated species ($25.7\%$ of total species) belonging to $20$ different families dominated by \textit{Polygonaceae}, \textit{Rosaceae}, \textit{Solanaceae}, \textit{Araceae}, \textit{Fabaceae}. They do not share any emitted VOC with other plants, since they do not release any protonated mass at all. That result has to be interpreted taking into account $G^P_2(V_2,E_2)$ construction procedure. Just the emissions exceeding the $Q_{\frac{3}{4}}$ of the corresponding protonated mass distribution were considered as relevant. In that sense that nodes are isolated from the rest of the graph and they do not emit VOCs.
\end{itemize}

Previous results are summarized in Tab.~\ref{tab:g11_famiglie}, which shows the dominant families in each cluster and how many species belong to that families. The list of species present in each cluster is reported in Tab.~\ref{tab:g11_species}. 

Cluster $2$, besides being the biggest one, is made up by those species corresponding to the highest weighted degree vertices in $G^P_2(V_2,E_2)$. That species work as highly connected nodes, and they share several VOCs with the other neighboring nodes. They correspond to the biggest yellow nodes in Fig.~\ref{fig:g11delQ3}. Here we list the principal ones: \textit{Lavandula spica} L. (Lavander), \textit{Foeniculum vulgare} Mill. (Fennel), \textit{Crithmum maritimum} L. (Samphire), \textit{Liquidambar styraciflua} L. (Sweetgum), \textit{Chrysanthemum indicum} L. (Chrisanth), \textit{Santolina chamaecyparissus} L. (Cotton lavender), \textit{Curcuma longa} L. (Turmeric), \textit{Cupressus sempervirens} L. (Mediterranean cypress), \textit{Ocimum basilicum} L. (Basil), \textit{Citrus x Aurantium} L. (Bitter orange), \textit{Tetradenia riparia} (Hochst.) Codd. (Ginger bush), \textit{Juniperus communis} L. (Juniper), \textit{Artemisia vulgaris} L. (Mugwort), \textit{Citrus x Limon} L. (Lemon), \textit{Stevia rebaudiana} (Stevia), \textit{Eucalyptus globulus} L. (Eucalyptus), \textit{Quercus ilex} L. (Holm oak), \textit{Hedera helix} L. (Ivy).  

Other species with as well a huge emission of VOCs are present in cluster $1$: \textit{Gossypium herbaceum} L. (Cotton), \textit{Plantago lanceolata} L. (Plantain), and \textit{Inula viscosa} L. (Inula) are the most connected aqua nodes in Fig.~\ref{fig:g11delQ3}.

Cluster $3$ (violet vertices in Fig.~\ref{fig:g11delQ3}.) turns out to be the most homogeneous one in terms of families composition, since it groups species belonging mainly to $Brassicaceae$ family,   characterized by the predominant emission of sulphur compounds.

\begin{table}[]
\centering
\resizebox{0.8\textwidth}{!}{
\begin{tabular}{|c|c|c|c|c|c|c|c|}
\hline
\hline
\multicolumn{2}{|c}{cluster $1$} & \multicolumn{2}{|c}{cluster $2$} & \multicolumn{2}{|c}{cluster $3$} & \multicolumn{2}{|c|}{cluster $4$} \\
\hline
\hline 
Rosaceae       & 4 & Asteraceae       & 5 & Brassicaceae  & 4 & Polygonaceae  & 3 \\
Asteraceae     & 3 & Apiaceae         & 3 & Actinidiaceae & 1 & Rosaceae      & 3 \\
Fabaceae       & 3 & Cannabaceae      & 3 & Fabaceae      & 1 & Solanaceae    & 3 \\
Ebenaceae      & 2 & Lamiaceae        & 3 &               &   & Araceae       & 2 \\
Plantaginaceae & 2 & Cupressaceae     & 2 &               &   & Fabaceae      & 2 \\
Solanaceae     & 2 & Magnoliaceae     & 2 &               &   & Apocynaceae   & 1 \\
Amaranthaceae  & 1 & Martyniaceae     & 2 &               &   & Aquifoliaceae & 1 \\
Asparagaceae   & 1 & Myrtaceae        & 2 &               &   & Asteraceae    & 1 \\
Betulaceae     & 1 & Rutaceae         & 2 &               &   & Crassulaceae  & 1 \\
Cyperaceae     & 1 & Sapindaceae      & 2 &               &   & Faboideae     & 1 \\
Ericaceae      & 1 & Solanaceae       & 2 &               &   & Hydrangeaceae & 1 \\
Fagaceae       & 1 & Araliaceae       & 1 &               &   & Iridaceae     & 1 \\
Iridoideae     & 1 & Brassicaceae     & 1 &               &   & Lauraceae     & 1 \\
Malvaceae      & 1 & Calycanthaceae   & 1 &               &   & Lythraceae    & 1 \\
Moraceae       & 1 & Caricaceae       & 1 &               &   & Malvaceae     & 1 \\
Oleaceae       & 1 & Composite        & 1 &               &   & Moraceae      & 1 \\
Paulowniaceae  & 1 & Convolvulaceae   & 1 &               &   & Oleaceae      & 1 \\
Platanaceae    & 1 & Fagaceae         & 1 &               &   & Poaceae       & 1 \\
Rhamnaceae     & 1 & Hamamelidaceae   & 1 &               &   & Portulacaceae & 1 \\
Salicaceae     & 1 & Lauraceae        & 1 &               &   & Vitaceae      & 1 \\
Sapindaceae    & 1 & Pinaceae         & 1 &               &   &               &   \\
               &   & Polygonaceae     & 1 &               &   &               &   \\
               &   & Rutacee          & 1 &               &   &               &   \\
               &   & Turneraceae      & 1 &               &   &               &   \\
               &   & Urticaceae       & 1 &               &   &               &   \\
               &   & Xanthorrhoeaceae & 1 &               &   &               &   \\
               &   & Zingiberaceae    & 1 &               &   &               &   \\
\hline
\hline
\textbf{Fam.} & \textbf{$\#$ Spec.} & \textbf{Fam.} &  \textbf{$\#$ Spec.} &\textbf{Fam.} & \textbf{$\#$ Spec.} & \textbf{Fam.} & \textbf{$\#$ Spec.} \\
\hline
\hline
\end{tabular}}
\caption{Plants families composition in each community extracted from third-quartile-based graph $G^P_2(V_2,E_2)$ by modularity (BL) algorithm, and the corresponding amount of species belonging to that families for each community. Exception made for cluster $3$ (violet), a huge families heterogeneity characterizes all the other communities.}
\label{tab:g11_famiglie}
\end{table}

\begin{table}[]
\centering
\resizebox{\textwidth}{!}{
\begin{tabular}{|c|c|c|c|c|c|c|c|}
\hline
\hline
\multicolumn{2}{|c}{cluster $1$} & \multicolumn{2}{|c}{cluster $2$} & \multicolumn{2}{|c}{cluster $3$} & \multicolumn{2}{|c|}{cluster $4$} \\
\hline
\hline
\textit{Mimosa pudica} L.                 & Fabaceae       & \textit{Ocimum basilicum} L.                          & Lamiaceae        & \textit{Brassica rapa} L.                       & Brassicaceae  & \textit{Zamioculcas zamiifolia} (Lodd.) & Araceae       \\
\textit{Cyperus papyrus} L.               & Cyperaceae     & \textit{Brassica oleracea} L. var  botrytis           & Brassicaceae     & \textit{Brassica oleracea} L. var acephala      & Brassicaceae  & \textit{Rheum rhabarbarum} L.           & Polygonaceae  \\
\textit{Ziziphus jujuba} Mill.            & Rhamnaceae     & \textit{Stevia rebaudiana}                            & Asteraceae       & \textit{Actinidia arguta} (Siebold \& Zucc.)    & Actinidiaceae & \textit{Hydrangea macrophylla} (Lam.)   & Hydrangeaceae \\
\textit{Platanus x Acerifolia} (Willd.)   & Platanaceae    & \textit{Cannabis sativa} L.                           & Cannabaceae      & \textit{Eruca sativa} (Mill.)                   & Brassicaceae  & \textit{Solanum marginatum} L.          & Solanaceae    \\
\textit{Plantago lanceolata} L.           & Plantaginaceae & \textit{Nicotiana tabacum} L.                         & Solanaceae       & \textit{Vicia faba} L.                          & Fabaceae      & \textit{Persea americana} Mill.         & Lauraceae     \\
\textit{Arbutus unedo} L.                 & Ericaceae      & \textit{Eucalyptus globulus} L.                       & Myrtaceae        & \textit{Lunaria annua} L.                       & Brassicaceae  & \textit{Vitis vinifera} L.              & Vitaceae      \\
\textit{Cotoneaster horizontalis} Decne.  & Rosaceae       & \textit{Ibicella lutea} L.                            & Martyniaceae     &                                        &               & \textit{Echeveria elegans} (Rose)       & Crassulaceae  \\
\textit{Sonchus oleraceus} L.             & Asteraceae     & \textit{Proboscidea parviflora} (Woot. \& Standl.)    & Martyniaceae     &                                        &               & \textit{Arundo donax} L.                & Poaceae       \\
\textit{Inula viscosa} L.                 & Asteraceae     & \textit{Quercus ilex} L.                              & Fagaceae         &                                        &               & \textit{Rumex acetosella} L.            & Polygonaceae  \\
\textit{Corylus avellana} L.              & Betulaceae     & \textit{Artemisia dracunculus} L.                     & Composite        &                                        &               & \textit{Acacia dealbata} Link           & Fabaceae      \\
\textit{Prunus armeniaca} L.              & Rosaceae       & \textit{Convolvulus cneorum} L.                       & Convolvulaceae   &                                        &               & \textit{Robinia pseudoacacia} L.        & Faboideae     \\
\textit{Acer campestre} L.                & Sapindaceae    & \textit{Juniperus communis} L.                        & Cupressaceae     &                                        &               & \textit{Olea europaea} L.               & Oleaceae      \\
\textit{Osmanthus heterophyllus} (G. Don) & Oleaceae       & \textit{Santolina chamaecyparissus} L.                & Asteraceae       &                                        &               & \textit{Fragaria vesca} L.              & Rosaceae      \\
\textit{Diospyros kaki} L.                & Ebenaceae      & \textit{Apium graveolens} L.                          & Apiaceae         &                                        &               & \textit{Rosa chinensis} (Jacq.)         & Rosaceae      \\
\textit{Ficus benjamina} L.               & Moraceae       & \textit{Ruta graveolens} L.                           & Rutacee          &                                        &               & \textit{Trifolium pratense} L.          & Fabaceae      \\
\textit{Populus alba} L.                  & Salicaceae     & \textit{Parietaria judaica} L.                        & Urticaceae       &                                        &               & \textit{Anthurium andreanum} Lind.      & Araceae       \\
\textit{Iris germanica} L.                & Iridoideae     & \textit{Cupressus sempervirens} L.                    & Cupressaceae     &                                        &               & \textit{Ficus carica} L.                & Moraceae      \\
\textit{Quercus cerris} L.                & Fagaceae       & \textit{Calycanthus floridus} L.                      & Calycanthaceae   &                                        &               & \textit{Ilex aquifolium} L.             & Aquifoliaceae \\
\textit{Tamarindus indica} L.             & Fabaceae       & \textit{Picea abies} L.                               & Pinaceae         &                                        &               & \textit{Pyrus communis} L.              & Rosaceae      \\
\textit{Salicornia europaea} L.           & Amaranthaceae  & \textit{Humulus lupulus} L. var. Cascade              & Cannabaceae      &                                        &               & \textit{Silybum marianum} L.            & Asteraceae    \\
\textit{Solanum quitoense} Lam.           & Solanaceae     & \textit{Humulus lupulus} L.                           & Cannabaceae      &                                        &               & \textit{Portulaca oleracea} L.          & Portulacaceae \\
\textit{Sansevieria trifasciata} Prain.   & Asparagaceae   & \textit{Hedera helix} L.                              & Araliaceae       &                                        &               & \textit{Capsicum chacoense} Hunz.       & Solanaceae    \\
\textit{Diospyros lotus} L.               & Ebenaceae      & \textit{Cardiospermum halicacabum} L.                 & Sapindaceae      &                                        &               & \textit{Withania somnifera} L.          & Solanaceae    \\
\textit{Linaria vulgaris} Mill.           & Plantaginaceae & \textit{Curcuma longa} L.                             & Zingiberaceae    &                                        &               & \textit{Rumex acetosa} L.               & Polygonaceae  \\
\textit{Taraxacum officinale} F.H. Wigg   & Asteraceae     & \textit{Foeniculum vulgare} Mill.                     & Apiaceae         &                                        &               & \textit{Punica granatum} L.             & Lythraceae    \\
\textit{Wisteria floribunda} (Willd.)     & Fabaceae       & \textit{Laurus nobilis} L.                            & Lauraceae        &                                        &               & \textit{Nerium oleander} L.             & Apocynaceae   \\
\textit{Gossypium herbaceum} L.           & Malvaceae      & \textit{Magnolia grandiflora} L.                      & Magnoliaceae     &                                        &               & \textit{Iris pallida} Lamm.             & Iridaceae     \\
\textit{Mespilus germanica} L.            & Rosaceae       & \textit{Citrus x Aurantium} L.                        & Rutaceae         &                                        &               & \textit{Hibiscus syriacus} L.           & Malvaceae     \\
\textit{Prunus persica} L.                & Rosaceae       & \textit{Carica papaya} L.                             & Caricaceae       &                                        &               &                                &               \\
\textit{Paulownia tomentosa} Steud.       & Paulowniaceae  & \textit{Aloe vera} L.                                 & Xanthorrhoeaceae &                                        &               &                                &               \\
\textit{Capsicum annuum} L.               & Solanaceae     & \textit{Liquidambar styraciflua} L.                   & Hamamelidaceae   &                                        &               &                                &               \\
                                 &                & \textit{Artemisia vulgaris} L.                        & Asteraceae       &                                        &               &                                &               \\
                                 &                & \textit{Magnolia liliiflora} (Desr.)                  & Magnoliaceae     &                                        &               &                                &               \\
                                 &                & \textit{Solanum lycopersicum} L.                      & Solanaceae       &                                        &               &                                &               \\
                                 &                & \textit{Chrysanthemum indicum} L.                     & Asteraceae       &                                        &               &                                &               \\
                                 &                & \textit{Tetradenia riparia} (Hochst.) Codd.           & Lamiaceae        &                                        &               &                                &               \\
                                 &                & \textit{Crithmum maritimum} L.                        & Apiaceae         &                                        &               &                                &               \\
                                 &                & \textit{Paullinia cupana} Kunth.                      & Sapindaceae      &                                        &               &                                &               \\
                                 &                & \textit{Coccoloba uvifera} L.                         & Polygonaceae     &                                        &               &                                &               \\
                                 &                & \textit{Turnera afrodisiaca} Ward.                    & Turneraceae      &                                        &               &                                &               \\
                                 &                & \textit{Citrus Limon} L.                              & Rutaceae         &                                        &               &                                &               \\
                                 &                & \textit{Psidium guajava} L.                           & Myrtaceae        &                                        &               &                                &               \\
                                 &                & \textit{Lavandula spica} L.                           & Lamiaceae        &                                        &               &                                &               \\
                                 &                & \textit{Dahlia pinnata} Cav.                          & Asteraceae       &                                        &               &                                &               \\
\hline
\hline                                 
\textbf{Spec.}                   & \textbf{Fam.}  & \textbf{Spec.}                              &  \textbf{Fam}     & \textbf{Spec.}                         &\textbf{Fam.}  & \textbf{Spec.}                 & \textbf{Fam.} \\           
\hline
\hline                                 
\end{tabular}}
\caption{Species composition of each cluster found in $G^P_2(V_2,E_2)$ third-quartile-based graph. Cluster $4$ is made up by the isolated nodes, i.e. by all that species which don't share any VOCs with all the other species.}
\label{tab:g11_species}
\end{table}

\subsubsection*{Selected VOCs graph}
Communities detection algorithms were applied to the $G^P_3(V_3, E_3)$ following the same procedure described for  $G^P_2(V_2, E_2)$ graph. The VOCs reduction reflected into a more clear picture of species reciprocal behaviour in terms of emitted protonated masses.
Besides the set of $28$ isolated nodes, tree big communities were detected. 

\begin{figure}
\centering
\includegraphics[width=1\textwidth]{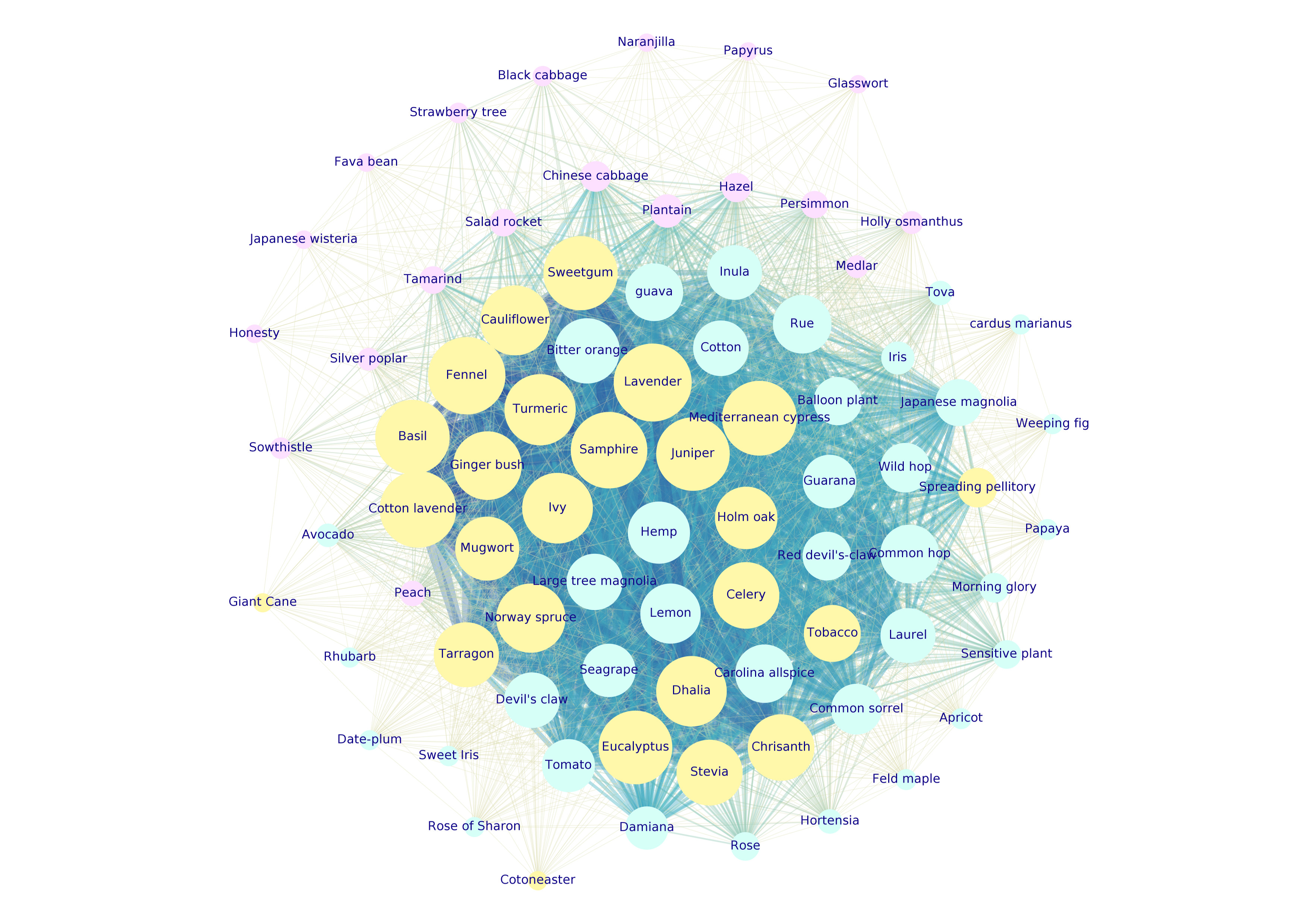}
\caption{$G^P_3(V_3,E_3)$ selected-VOCs graph. Each color corresponds to one detected community: cluster $1$ (aqua), cluster $2$ (yellow), cluster $3$ (violet). The $28$ isolated nodes are not shown (cluster $4$). Nodes dimension is proportional to their weighted degree. Edges thickness is proportional to the edge weight. Still \textit{Lavandula spica} L. (Lavender), \textit{Foeniculum vulgare} Mill. (Fennel), \textit{Santolina chamaecyparissus} L. (Cotton lavender), and \textit{Crithmum maritimum} L. (Samphire) are some of the most VOCs emitting species.} 
\label{fig:g13}
\end{figure}

Figure~\ref{fig:g13} shows $G^P_3(V_3, E_3)$ graph partitioning. The graph's nodes are coloured according to their community membership. Such as for $G^P_2(V_2, E_2)$ bipartite projection graph, the biggest nodes correspond to those species which share several VOCs with the other neighboring species. Analogously, edges weights are proportional to the amount of VOCs shared by each couple of adjacent vertices. 

Again cluster $2$ (yellow nodes, Fig.~\ref{fig:g13}) is made up by the highest-weighted-degree nodes. In other terms the species corresponding to yellow nodes are the most interconnected ones: \textit{Lavandula spica} L. (Lavender), \textit{Foeniculum vulgare} Mill. (Fennel), \textit{Santolina chamaecyparissus} L. (Cotton lavender), \textit{Crithmum maritimum} L. (Samphire), \textit{Cupressus sempervirens} L. (Mediterranean cypress), \textit{Ocimum basilicum} L. (Basil), \textit{Liquidambar styraciflua} L. (Sweetgum), \textit{Eucalyptus globulus} L. (Eucalyptus), \textit{Juniperus communis} L. (Juniper), \textit{Curcuma longa} L. (Turmeric), \textit{Hedera helix} L. (Ivy), \textit{Dahlia pinnata} Cav. (Dhalia), \textit{Brassica oleracea} L. \textit{var botrytis} (Cauliflower), \textit{Picea abies} L. (Norway spruce), \textit{Tetradenia riparia (Hochst.)} Codd. (Ginger bush), \textit{Apium graveolens} L. (Celery), \textit{Stevia rebaudiana} (Stevia), \textit{Artemisia dracunculus} L. (Tarragon), \textit{Artemisia vulgaris} L. (Mugwort), \textit{Quercus ilex} L. (Holm oak). That result is fully in agreement with the previous one. 

Some highly connected nodes are also present in cluster $1$ (aqua nodes, Fig.~\ref{fig:g13}), such as for example: \textit{Citrus x Aurantium} L. (Bitter orange), \textit{Cannabis sativa} L. (Hemp), \textit{Citrus x Limon} L. (Lemon), \textit{Humulus lupulus} L. \textit{var. Cascade} (Common hop), \textit{Ruta graveolens} L. (Rue), \textit{Calycanthus floridus} L. (Carolina allspice) and \textit{Psidium guajava} L. (guava).

Cluster $3$ is still homogeneously made-up by $Brassicaceae$ species (violet vertex in Fig.~\ref{fig:g13}).

Hereafter the four communities are described in term of dominating families and clustering protonated masses. 
\begin{itemize}
\item{cluster $1$}: it is the biggest community, made up by $37$ species ($33.9\%$ of the total species amount) grouped into $23$ families; dominant families: \textit{Cannabaceae}, \textit{Polygonaceae}, \textit{Sapindaceae}, \textit{Asteraceae}, \textit{Lauraceae}, \textit{Magnoliaceae}, \textit{Malvaceae}, \textit{Martyniaceae}, \textit{Rosaceae}, and \textit{Solanaceae}. This community is characterized by an high heterogeneity in terms of its families composition. The species belonging to that cluster release in particular PM$93$ (Tp-f, $22$ species), PM$109$ (Tp-f) and PM$137$ (Tp/STp-f) ($26$ species), PM$95$ (STp-f), PM$121$ (Tp-f), PM$123$ (Tp/STp-f), PM$149$ (Tp/STp-f), PM$205$ (STp)(more than $20$ species). The $m/z$ listed above probably refer to terpenes compounds and almost all of them are found in plant belonging to cluster $2$ of the previous analysis. Indeed, the actual cluster $1$ shares with the previous cluster $2$ more than 51\% of plant species (Tab.~\ref{tab:g11_species} and Tab.~\ref{tab:g13_species}), including \textit{Citrus} spp. In this community the species that release sulfur compounds (PM$49$ and PM$63$) are also found, such as: \textit{Ruta graveolens} L. (Rue), \textit{Inula viscosa} L. (Inula), \textit{Psidium guajava} L. (Guava), \textit{Gossypium herbaceum} L. (Cotton), and \textit{Citrus x Aurantium} L. (Bitter orange), which together with \textit{Cannabis sativa} L. (Hemp), and \textit{Citrus x Limon} L. (Lemon) are among the most emitting species. Interestingly, species from \textit{Brassicaceae} family, typically rich in sulfur compounds\cite{geervliet1997comparative}, are not included in this cluster.

\item{cluster $2$}: $25$ species ($22.9\%$ of database total species) grouped in $16$ families; prevailing families: \textit{Asteraceae} ($5$ species), \textit{Apiaceae}, \textit{Lamiaceae}, and \textit{Cupressaceae}. This community is made up by those species which are the most active in terms of VOCs emission, in agreement with the species gathered in cluster $2$ of the previous analysis; see yellow nodes in Fig.~\ref{fig:g11delQ3} and Tab.~\ref{tab:g11_species}. As an example, we just list the most interconnected nodes: \textit{Lavandula spica} L. (Lavender), \textit{Foeniculum vulgare} Mill. (Fennel), \textit{Santolina chamaecyparissus} L. (Cotton lavender, known for its smell), \textit{Crithmum maritimum} L. (Samphire) (found in cluster $2$ of the previous analysis). Cauliflower is also found here. Again, an high heterogeneity characterizes the families distribution. Accordingly, the species belonging to this cluster release some volatiles already highlighted for the previous cluster $2$; in fact, the most released VOC is PM$153$ (Tp), emitted by $24$ species, followed by PM$93$ (Tp-f), PM$95$ (STp-f), PM$121$ (Tp-f), PM$123$ (Tp/STp-f), PM$149$ (Tp/STp-f) (released by $23$ species), and PM$109$ (Tp-f), PM$119$ (Tp-f), PM$133$ (Tp), PM$137$ (Tp/STp-f), PM$143$ (K/Ald), PM$151$ (Tp/Tp-f), PM$205$ (STp)(emitted by more than $20$ species). Except for the ketone PM$143$, they are all terpenes compounds.

\item{cluster $3$}: $19$ species ($17.5\%$ of total species) from $13$ families only: \textit{Brassicaceae}, \textit{Actinidiaceae}, and \textit{Fabaceae}. All these species emit in particular sulphur compounds PM$49$ (SC) and PM$63$ (SC) ($13$ and $12$ species, respectively), while just few of them also release PM$93$, PM$95$, and PM$153$ (Tp-f, STp-f and Tp, respectively). \textit{Brassica rapa} L. (Chinese cabbage) species again distinguishes, being the only one which emits PM$143$ (K/Ald). This cluster is the most stable and it corresponds to cluster $3$ of the previous analysis. It shows an homogenous families composition, since it groups all the \textit{Brassicaceae} species, exception made for the \textit{Brassica oleracea} L. \textit{var botrytis} (Cauliflower) species, in agreement with previous analysis. 

\item{cluster $4$}: $28$ isolated species ($25.7\%$ of total species) belonging to $23$ different families dominated by \textit{Solanaceae}, \textit{Araceae}, \textit{Fabaceae}, \textit{Rosaceae}. As for the previous analysis on graph $G^P_2(V_2, E_2)$ the isolated nodes correspond to species which do not emit any VOCs. 
\end{itemize}

A detailed description of the plants families and species composition of each cluster of $G^P_3(V_3, E_3)$ graph is provided in Tab.~\ref{tab:g13_families} and Tab.~\ref{tab:g13_species}, respectively.

\begin{table}[]
\centering
\resizebox{0.8\textwidth}{!}{
\begin{tabular}{|c|c|c|c|c|c|c|c|}
\hline
\hline
\multicolumn{2}{|c}{cluster $1$} & \multicolumn{2}{|c}{cluster $2$} & \multicolumn{2}{|c}{cluster $3$} & \multicolumn{2}{|c|}{cluster $4$} \\
\hline
\hline
Cannabaceae    & 3 & Asteraceae     & 5 & Brassicaceae   & 4 & Solanaceae       & 3 \\
Polygonaceae   & 3 & Apiaceae       & 3 & Fabaceae       & 3 & Araceae          & 2 \\
Sapindaceae    & 3 & Lamiaceae      & 3 & Rosaceae       & 2 & Fabaceae         & 2 \\
Asteraceae     & 2 & Cupressaceae   & 2 & Amaranthaceae  & 1 & Rosaceae         & 2 \\
Lauraceae      & 2 & Araliaceae     & 1 & Asteraceae     & 1 & Actinidiaceae    & 1 \\
Magnoliaceae   & 2 & Brassicaceae   & 1 & Betulaceae     & 1 & Apocynaceae      & 1 \\
Malvaceae      & 2 & Composite      & 1 & Cyperaceae     & 1 & Aquifoliaceae    & 1 \\
Martyniaceae   & 2 & Fagaceae       & 1 & Ebenaceae      & 1 & Asparagaceae     & 1 \\
Rosaceae       & 2 & Hamamelidaceae & 1 & Ericaceae      & 1 & Asteraceae       & 1 \\
Rutacee        & 2 & Myrtaceae      & 1 & Oleaceae       & 1 & Crassulaceae     & 1 \\
Solanaceae     & 2 & Pinaceae       & 1 & Plantaginaceae & 1 & Faboideae        & 1 \\
Calycanthaceae & 1 & Poaceae        & 1 & Salicaceae     & 1 & Fagaceae         & 1 \\
Caricaceae     & 1 & Rosaceae       & 1 & Solanaceae     & 1 & Lythraceae       & 1 \\
Convolvulaceae & 1 & Solanaceae     & 1 &                &   & Moraceae         & 1 \\
Ebenaceae      & 1 & Urticaceae     & 1 &                &   & Oleaceae         & 1 \\
Fabaceae       & 1 & Zingiberaceae  & 1 &                &   & Paulowniaceae    & 1 \\
Hydrangeaceae  & 1 &                &   &                &   & Plantaginaceae   & 1 \\
Iridaceae      & 1 &                &   &                &   & Platanaceae      & 1 \\
Iridoideae     & 1 &                &   &                &   & Polygonaceae     & 1 \\
Moraceae       & 1 &                &   &                &   & Portulacaceae    & 1 \\
Myrtaceae      & 1 &                &   &                &   & Rhamnaceae       & 1 \\
Turneraceae    & 1 &                &   &                &   & Vitaceae         & 1 \\
               &   &                &   &                &   & Xanthorrhoeaceae & 1 \\
\hline
\hline
\textbf{Fam.} & \textbf{$\#$ Spec.} & \textbf{Fam.} &  \textbf{$\#$ Spec.} &\textbf{Fam.} & \textbf{$\#$ Spec.} & \textbf{Fam.} & \textbf{$\#$ Spec.} \\
\hline
\hline
\end{tabular}}
\caption{Plants families composition in each community extracted from graph $G^P_3(V_3,E_3)$ by modularity (BL) algorithm, and the corresponding amount of species belonging to that families for each community.}
\label{tab:g13_families}
\end{table}

\begin{table}[]
\centering
\resizebox{\textwidth}{!}{%
\begin{tabular}{|c|c|c|c|c|c|c|c|}
\hline
\hline
\multicolumn{2}{|c}{cluster $1$} & \multicolumn{2}{|c}{cluster $2$} & \multicolumn{2}{|c}{cluster $3$} & \multicolumn{2}{|c|}{cluster $4$} \\
\hline
\hline
\textit{Cannabis sativa} L.                        & Cannabaceae    & \textit{Ocimum basilicum} L.                & Lamiaceae      & \textit{Brassica rapa} L.                  & Brassicaceae   & \textit{Zamioculcas zamiifolia} (Lodd.)         & Araceae          \\
\textit{Ibicella lutea} L.                         & Martyniaceae   & \textit{Brassica oleracea} L. var  botrytis & Brassicaceae   & \textit{Cyperus papyrus} L.                & Cyperaceae     & \textit{Solanum marginatum} L.                  & Solanaceae       \\
\textit{Proboscidea parviflora} (Woot. Et Standl.) & Martyniaceae   & \textit{Stevia rebaudiana}                  & Asteraceae     & \textit{Brassica oleracea} L. var acephala & Brassicaceae   & \textit{Vitis vinifera} L.                      & Vitaceae         \\
\textit{Convolvulus cneorum} L.                    & Convolvulaceae & \textit{Nicotiana tabacum} L.               & Solanaceae     & \textit{Plantago lanceolata} L.            & Plantaginaceae & \textit{Echeveria elegans} (Rose)               & Crassulaceae     \\
\textit{Rheum rhabarbarum} L.                      & Polygonaceae   & \textit{Eucalyptus globulus} L.             & Myrtaceae      & \textit{Arbutus unedo} L.                  & Ericaceae      & \textit{Rumex acetosella} L.                    & Polygonaceae     \\
\textit{Ruta graveolens} L.                        & Rutacee        & \textit{Quercus ilex} L.                    & Fagaceae       & \textit{Sonchus oleraceus} L.              & Asteraceae     & \textit{Acacia dealbata} Link                   & Fabaceae         \\
\textit{Hydrangea macrophylla} (Lam.)              & Hydrangeaceae  & \textit{Artemisia dracunculus} L.           & Composite      & \textit{Corylus avellana} L.               & Betulaceae     & \textit{Ziziphus jujuba} Mill.                  & Rhamnaceae       \\
\textit{Persea americana} Mill.                    & Lauraceae      & \textit{Juniperus communis} L.              & Cupressaceae   & \textit{Osmanthus heterophyllus} (G. Don)  & Oleaceae       & \textit{Robinia pseudoacacia} L.                & Faboideae        \\
\textit{Mimosa pudica} L.                          & Fabaceae       & \textit{Santolina chamaecyparissus} L.      & Asteraceae     & \textit{Diospyros kaki} L.                 & Ebenaceae      & \textit{Olea europaea} L.                       & Oleaceae         \\
\textit{Humulus lupulus} L. var. Cascade           & Cannabaceae    & \textit{Arundo donax} L.                    & Poaceae        & \textit{Populus alba} L.                   & Salicaceae     & \textit{Trifolium pratense} L.                  & Fabaceae         \\
\textit{Humulus lupulus} L.                        & Cannabaceae    & \textit{Parietaria judaica} L.              & Urticaceae     & \textit{Tamarindus indica} L.              & Fabaceae       & Anthurium andreanum Lind.              & Araceae          \\
\textit{Rosa chinensis} (Jacq.)                    & Rosaceae       & \textit{Cupressus sempervirens} L.          & Cupressaceae   & \textit{Salicornia europaea} L.            & Amaranthaceae  & \textit{Ficus carica} L.                        & Moraceae         \\
\textit{Cardiospermum halicacabum} L.              & Sapindaceae    & \textit{Picea abies} L.                     & Pinaceae       & \textit{Eruca sativa} (Mill.)              & Brassicaceae   & \textit{Ilex aquifolium} L.                     & Aquifoliaceae    \\
\textit{Silybum marianum} L.                       & Asteraceae     & \textit{Hedera helix} L.                    & Araliaceae     & \textit{Solanum quitoense} Lam.            & Solanaceae     & \textit{Platanus x Acerifolia} (Willd.)         & Platanaceae      \\
\textit{Laurus nobilis} L.                         & Lauraceae      & \textit{Curcuma longa} L.                   & Zingiberaceae  & \textit{Vicia faba} L.                     & Fabaceae       & \textit{Pyrus communis} L.                      & Rosaceae         \\
\textit{Inula viscosa} L.                          & Asteraceae     & \textit{Foeniculum vulgare} Mill.           & Apiaceae       & \textit{Wisteria floribunda} (Willd.)      & Fabaceae       & \textit{Actinidia arguta} (Siebold \& Zucc.) & Actinidiaceae    \\
\textit{Magnolia grandiflora} L.                   & Magnoliaceae   & \textit{Cotoneaster horizontalis} Decne.    & Rosaceae       & \textit{Mespilus germanica} L.             & Rosaceae       & \textit{Portulaca oleracea} L.                  & Portulacaceae    \\
\textit{Prunus armeniaca} L.                       & Rosaceae       & \textit{Liquidambar styraciflua} L.         & Hamamelidaceae & \textit{Prunus persica} L.                 & Rosaceae       & \textit{Aloe vera} L.                           & Xanthorrhoeaceae \\
\textit{Acer campestre} L.                         & Sapindaceae    & \textit{Artemisia vulgaris} L.              & Asteraceae     & \textit{Lunaria annua} L.                  & Brassicaceae   & \textit{Quercus cerris} L.                      & Fagaceae         \\
\textit{Calycanthus floridus} L.                   & Calycanthaceae & \textit{Apium graveolens} L.                & Apiaceae       &                                   &                & \textit{Fragaria vesca} L.                      & Rosaceae         \\
\textit{Carica papaya} L.                          & Caricaceae     & \textit{Chrysanthemum indicum} L.           & Asteraceae     &                                   &                & \textit{Withania somnifera} L.                  & Solanaceae       \\
\textit{Ficus benjamina} L.                        & Moraceae       & \textit{Tetradenia riparia} (Hochst.) Codd. & Lamiaceae      &                                   &                & \textit{Sansevieria trifasciata} Prain.         & Asparagaceae     \\
\textit{Iris germanica} L.                         & Iridoideae     & \textit{Crithmum maritimum} L.              & Apiaceae       &                                   &                & \textit{Linaria vulgaris} Mill.                 & Plantaginaceae   \\
\textit{Magnolia liliiflora} (Desr.)               & Magnoliaceae   & \textit{Lavandula spica} L.                 & Lamiaceae      &                                   &                & \textit{Taraxacum officinale} F.H. Wigg         & Asteraceae       \\
\textit{Capsicum chacoense} Hunz.                  & Solanaceae     & \textit{Dahlia pinnata} Cav.                & Asteraceae     &                                   &                & \textit{Punica granatum} L.                     & Lythraceae       \\
\textit{Solanum lycopersicum} L.                   & Solanaceae     &                                    &                &                                   &                & \textit{Nerium oleander} L.                     & Apocynaceae      \\
\textit{Paullinia cupana} Kunth.                   & Sapindaceae    &                                    &                &                                   &                & \textit{Paulownia tomentosa} Steud.             & Paulowniaceae    \\
\textit{Coccoloba uvifera} L.                      & Polygonaceae   &                                    &                &                                   &                & \textit{Capsicum annuum} L.                     & Solanaceae       \\
\textit{Turnera afrodisiaca} Ward.                 & Turneraceae    &                                    &                &                                   &                &                                        &                  \\
\textit{Diospyros lotus} L.                        & Ebenaceae      &                                    &                &                                   &                &                                        &                  \\
\textit{Citrus x Limon} L.                         & Rutaceae       &                                    &                &                                   &                &                                        &                  \\
\textit{Citrus x Aurantium} L.                      & Rutaceae       &                                    &                &                                   &                &                                        &                  \\
\textit{Psidium guajava} L.                        & Myrtaceae      &                                    &                &                                   &                &                                        &                  \\
\textit{Rumex acetosa} L.                          & Polygonaceae   &                                    &                &                                   &                &                                        &                  \\
\textit{Gossypium herbaceum} L.                    & Malvaceae      &                                    &                &                                   &                &                                        &                  \\
\textit{Iris pallida} Lamm.                        & Iridaceae      &                                    &                &                                   &                &                                        &                  \\
\textit{Hibiscus syriacus} L.                      & Malvaceae      &                                    &                &                                   &                &                                        &                  \\
\hline
\hline                                 
\textbf{Spec.}                   & \textbf{Fam.}  & \textbf{Spec.}                              &  \textbf{Fam}     & \textbf{Spec.}                         &\textbf{Fam.}  & \textbf{Spec.}                 & \textbf{Fam.} \\           
\hline
\hline                                 
\end{tabular}}
\caption{Species composition of each cluster found in $G^P_3(V_3,E_3)$ graph. Cluster $4$ is made up by the isolated nodes, i.e. all that species which do not emit any VOC.}
\label{tab:g13_species}
\end{table}

\subsubsection*{Features graph, $G_3^{PM}$}\label{sec:originalGraph}
\begin{figure}
\centering
\includegraphics[width=1\textwidth]{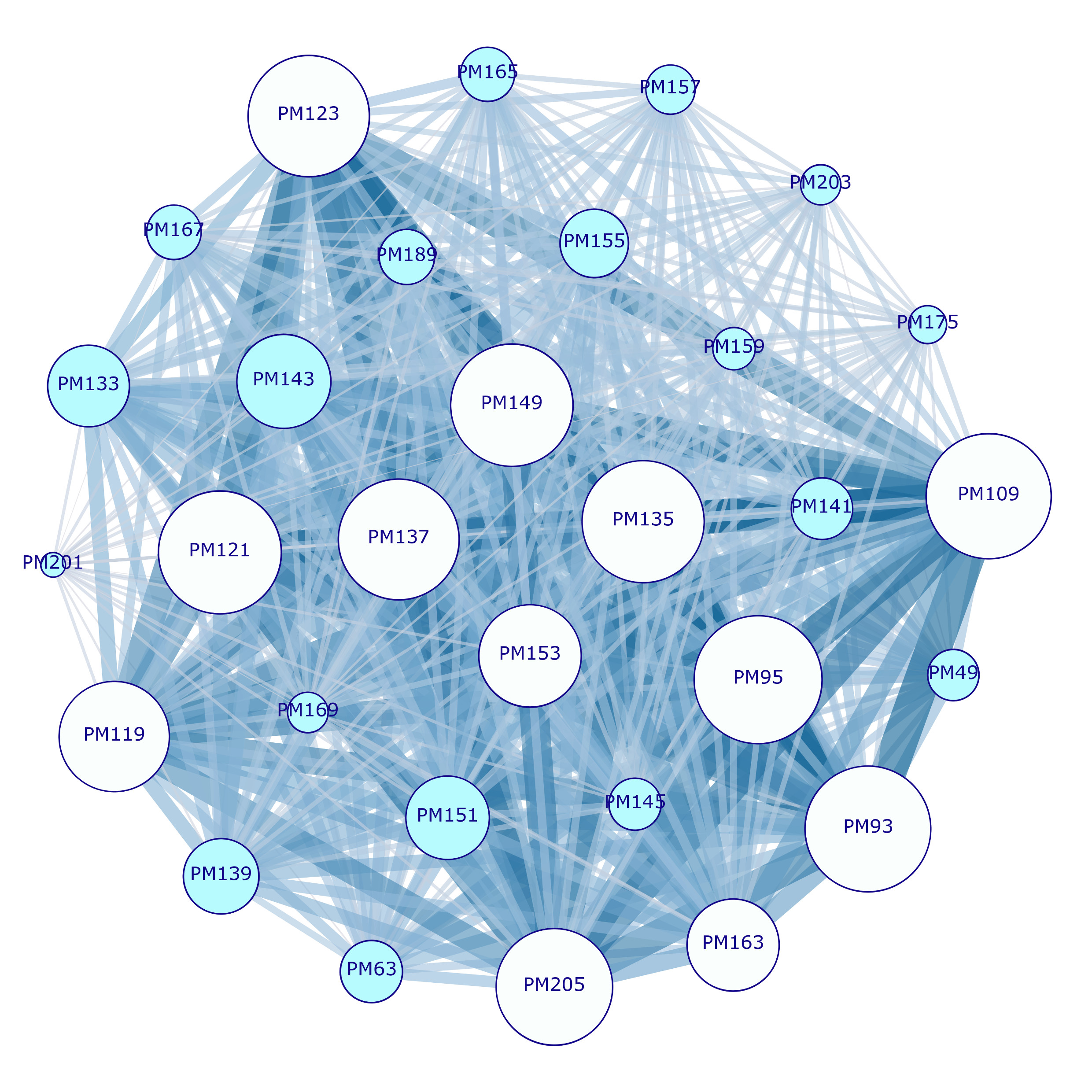}
\caption{$G^{PM}_3(V_{3b},E_{3b})$ features graph. Each vertex correspond to one VOC. Edges thickness is proportional to the amount of shared species. Nodes dimensions are proportional to their weighted degree. The most interconnected VOCs are evident (lighter color): PM$93$ (Tp-f), PM$95$ (STp-f), PM$109$ (Tp-f), PM$121$ (Tp-f), PM$123$ (Tp/STp-f), PM$135$ (Tp/STp-f), PM$137$ (Tp/STp-f), PM$149$ (Tp/STp-f), PM$205$ (STp). The protonated mass PM$201$ (Tp), on the contrary, is the less interconnected node, thus turning out to be the less commonly shared emitted VOC.} 
\label{fig:g13b}
\end{figure}

The second bipartite projection of graph $G_3(V,E)$, i.e. the graph of VOCs $G^{PM}_3(V_{3b},E_{3b})$ is shown in Fig.~\ref{fig:g13b}. The graph is made up by $V_{3b} = 30$ vertices (each corresponding to one protonated mass), and $E_{3b} = 435$ edges. Usually a bipartite graph is based on the representation of different individuals according to the common properties they share. Here the emitted VOCs are the analogous of features, since the most two plants emit the same volatiles the most they are similar. We chose to show only the results coming from the second bipartite projection of graph $G_3(V,E)$, since we obtained similar results for $G_2(V,E)$. Graph $G_1(V,E)$ is not considered since from the previous analyses it turned out to be less suitable to describe data as a network.
Colors here help the reader to distinguish between the most and less interconnected VOCs. Such as for the species-based bipartite projection graph, some protonated masses are highly connected with their neighborhoods. The highest value of weighted degree is recorded for PM$95$ ($s_{max} = 679$), followed by PM$93$, PM$109$, PM$121$, PM$149$, PM$135$, PM$123$, PM$137$, PM$205$ (light blue vertices in Fig.~\ref{fig:g13b}). All that VOCs are shared by a large number of species, and they are terpenes compounds; accordingly, they are the responsible for the species grouping in the first two communities of graph $G^{P}_{3}(V_{3},E_{3})$ (aqua and yellow clusters in Fig.~\ref{fig:g13}), made up by species rich in such types of compounds. Indeed, terpenes are the largest and assorted group of plant natural products, including hemiterpenes (C$_{5}$), monoterpenes (C$_{10}$), sesquiterpene (C$_{15}$), homoterpenes (C$_{11}$ and C$_{16}$), some diterpene (C$_{20}$) and triterpene (C$_{30}$), that are easily released into the atmosphere.
The highest amount of species shared between two VOCs is observed between all the following couples of VOCs: PM$93$--PM$95$ and PM$93$--PM$109$ (respectively $48$ and $46$ maximum numbers of species), PM$95$--PM$109$, PM$109$--PM$137$, PM$93$--PM$121$, PM$95$--PM$121$, PM$95$--PM$135$, PM$95$--PM$137$, PM$95$--PM$149$, PM$109$--PM$121$, PM$93$--PM$123$. Their corresponding links are the thickest ones (highest link weights) in Fig.~\ref{fig:g13b}. In most cases, plants share two compounds belonging to the same chemical class; for example, PM$95$--PM$109$ is a couple of sesquiterpenes and/or sesquiterpenes fragments, while PM$93$--PM$123$ are terpenes and/or terpenes fragments. It’s worth noting that sesquiterpenes have a distinct biochemical pathway from that of other hemiterpenes\cite{dudareva2013biosynthesis}, thus it is more expectable that a plant species emits, simultaneously, two or more VOCs of the same class instead of the combination of VOCs of different classes. However, terpenes biosynthesis is very complex\cite{sun2016my} and uses many separated pathways, and cases of plants producing isoprene (terpenes building unit) but not other monoterpenes (and viceversa) have been frequently reported\cite{lindfors2000biogenic}.

On the contrary, the two sulphur compounds PM$49$ and PM$63$, which considerably determine the assembling of the violet cluster in Fig.~\ref{fig:g13}, are small dimension nodes, since the species they share are homogeneous in term of family composition, but they are few. Among volatile organic sulfur compounds, dimethylsulfide (DMS, PM$63$) and methanethiol (MT, PM$49$) are two of the most frequent products of plant metabolism. Their biosynthetic pathways share the role of a common lyase enzyme (dimethylsulfoniopropionate, DMSP) that is not widely distributed in terrestrial plants\cite{bentley2004environmental}. 

Finally, PM$201$ (Tp), PM$169$ (aldehydes, Ald, a product of monoterpene oxidation), and PM$159$ (acids/esters, Ac/Es) are some of the less interconnected VOCS.

\section*{Conclusions}\label{sec:discussion}
Volatile organic compounds (VOCs), that represent a crucial component of a plant’s phenotype\cite{dicke2010induced}, have been analysed by bipartite networks methodology in order to classify plants species. In particular, several quantitative measures coming from Complex Network Theory\cite{ma2016wiener,li2013note,cao2014extremality} have been applied to uncover eventual similarities between the species in term of their VOCs emissions. To assure the reliability and robustness of the results, different classical and advanced community detection algorithms have been applied, and only the comparable results were retained. Moreover data have been pre-processed by means of both descriptive and quantitative statistical methods, to better focus on data behaviour. VOCs time series, obtained by recording the emissions content for each available species, suggest the presence of spike-like pulses (corresponding to few species), exceeding from a quite flat background signal. Each VOC turns out to be emitted by few species in a very large quantity, with respect to all the other species emissions of the same protonated mass. 

After a preliminary test performed on the whole dataset, some VOCs have been excluded. In fact, some volatiles, especially C6 compounds and acetaldehyde, can occur in response to external stress, including wounding; this should be taken into account when using these compounds for communities detection analysis. Using a reduced dataset, community detection suggested the presence of $4$ clusters. Two communities are made up by highly VOCs-emitting species. We recall here the most interconnected nodes: \textit{Lavandula spica} L. (Lavender), \textit{Foeniculum vulgare} Mill. (Fennel), \textit{Santolina chamaecyparissus} L. (Cotton lavender), \textit{Crithmum maritimum} L. (Samphire), \textit{Cupressus sempervirens} L. (Mediterranean cypress), \textit{Ocimum basilicum} L. (Basil) (for cluster $1$); \textit{Citrus x Aurantium} L.  (Bitter orange), \textit{Cannabis sativa} L. (Hemp), \textit{Citrus x Limon} L. (Lemon), \textit{Humulus lupulus} L. \textit{var. Cascade} (Common hop), \textit{Ruta graveolens} L. (Rue), \textit{Calycanthus floridus} L. (Carolina allspice) and \textit{Psidium guajava} L. (guava) (for cluster $2$).
A third community clearly groups species belonging to $Brassicaceae$ family, turning out to be quite homogeneous in terms of clusters families composition. Finally, a fourth community highlights all those species which, by network construction, are not sharing any VOCs emission with the other species. See previous Section ``Community detection analysis'' for more details.

The second bipartite projection confirmed terpenes compounds and sulphur compounds to be the two chemical classes most responsible for species classification. Indeed, the chemistry of volatiles has been shown to be species-specific\cite{dudareva2013biosynthesis}; for example, species characterized by terpenes and nitrogen-containing compounds as floral volatiles are different from species releasing sulphur-containing volatiles\cite{dobson2006relationship}. Moreover, terpenes compounds emitted by plant species (the so-called “terpenome”\cite{kumari2014essoildb}), are the major constituents of plants essential oils\cite{edris2007pharmaceutical}, and can be used to distinguish different species; in this study, although the exact chemical definition of the compounds involved is beyond the purpose, community detection highlighted two well defined groups (clusters $1$ and $2$) of species that emit different terpenes compounds.
In conclusion, complex network analysis confirms to be an advantageous methodology to uncover plants relationships also related to the way they react to the environment in which they live. That result strengthens previous findings obtained by applying Complex Network Theory to the plants morphological features\cite{vivaldo2016networks}. A similar approach can be extended to different fields in botanic framework, such as plant ecology, psychophysiology and plant communication.

\section*{Methods}\label{sec:methods}

\subsection*{Data}
PTR-ToF-MS has been used in this study as the detector for the organic compounds emitted by leaf samples. A full description of this tool, with its advantages and disadvantages, can be found elsewhere\cite{lindinger1998proton,jordan2009high,taiti2016assessing}. The compounds emitted by different leaves were transported from the air stream where collided with H$_3$O$^{+}$ reagent ion inside the drift tube.
The analysis was carried out as follows: each leaf samples was placed into $3/4$ L glass jar (Bormioli, Italy) provided of glass stopper fitted with two Teflon tubes connected respectively to the PTR-ToF-MS ($8000$, Ionicon Analitic GmbH, Innsbruck, Austria) and the zero air generator (Peak Scientific instruments, USA). Each sample was obtained by cutting pieces of representative mature and healthy leaves from three different plant exemplars ($5$ g total weight). For each plant species, three replicates (three different jar) were evaluated. An overview of the plants used is shown in Tab~\ref{tab:g11_species} and Tab~\ref{tab:g13_species}, for a total of $109$ species belonging to $56$ plant families. Before each leaf sample analysis, the glass jar was exposed to $1$ minute of purified air flux ($100$ sccm) to remove all the VOCs accumulated in the head space during the time between sample preparation; then, a blank air sample was taken and subsequently used for background correction. All measurements were conducted in an air-conditioned room, with temperature and humidity respectively set at $20 \pm 3 ^{\circ}$C and 65\% \cite{mancuso2015soil}, and using the same PTR-ToF-MS instrumental parameters: drift pressure = $2.30$ mbar, drift temperature = $60^{\circ}$C and inlet temperature = $40^{\circ}$C, drift voltage = $600$ V, extraction voltage at the end of the tube (Udx) $35$ V, which resulted in E/N ratio of $140$ Td ($1$ Td $= 10-17$ Vcm$^{-2}$). This setup allowed a good balance between excessive water cluster formation and product ion fragmentation\cite{pang2015biogenic}. Moreover, the inlet flux was set to $100$ sscm. The internal calibration of ToF spectra was based on m/z $= 29.997$ (NO$^+$), m/z $= 59.049$ (C$_3$H$_7$O$^+$) and m/z = $137.132$ (C$_{10}$H$_{17}^+$) and was performed off-line after dead time correction; for peak quantification, the resulting data were corrected according to the duty cycle. Data were recorded with the software TOF-DAQ (Tofwerk AG, Switzerland), the sampling time for each channel of TOF acquisition was 0.1 ns, acquiring 1 spectrum per second, for a mass spectrum range between m/z $20$ and m/z $220$. The raw data were normalized to the primary ion signal from counts per seconds (cps) to normalized counts per second (\textit{n}cps) as described by \textit{Herbig et al.}\cite{herbig2009line}. Data were filtered following the procedure used by \textit{Taiti et al.}\cite{taiti2016sometimes} and used for statistical analysis. In this manner, a dataset comprised of mean mass spectra for each sample analyzed was compiled. Finally, the tentative identifications of peaks was performed on the basis of an high mass resolution and rapid identification of compounds with a high level of confidence\cite{lanza2015selective}. Further characterization of VOCs belonging to certain chemical classes such as terpenes, which are prone to fragmentation, was attempted using literature data on fragmentation of standards during PTR-ToF-MS analysis \cite{maleknia2007ptr,kim2009measurement,demarcke2009laboratory}. Similar approach was performed for the other identified compound, e.g. following \textit{Papurello et al.}\cite{papurello2012monitoring} and \textit{Liu et al.}\cite{liu2013experimental} for sulfur compounds, \textit{Loreto et al.}\cite{loreto2006induction}, \textit{Brilli et al.}\cite{brilli2011detection}, \textit{Degen et al.}\cite{degen2004high}, and \textit{Wu et al.}\cite{wu2008comparison} for wounding-related VOCs, and \textit{Schwartz et al.}\cite{schwarz2009determining} and \textit{Soukoulis et al.}\cite{soukoulis2013ptr} for aldehydes, ketones and alcohols.

\subsection*{Descriptive statistics: boxplots}
Boxplots are an intuitive graphical non-parametric method particularly suitable to visualize the distribution of continuous univariate data, firstly proposed by \textit{Tukey}\cite{tukey1977exploratory}. None a-priory assumption is made on the underlying statistical distribution. Boxplots show information about data location and spread, by starting from the estimation of the second quartile (or median, $Q_2$) and of the interquartile range ($IQR$), where $IQR = Q_3 - Q_1$, and $Q_3$ and $Q_1$ are the third and first quartiles, respectively. Boxplots are also known as box-and-whisker plots. The rectangular box is related to the data quartiles, and, more in details, the left and right sides of the rectangle correspond respectively to $Q_1$ and $Q_3$. The whiskers are lines extending from the box till lower and upper first outliers. It follows that the boxplot width visually shows the sample $IQR$, the vertical band drawn inside the box represents the median, and as a whole the box is a measure of the data dispersion and skewness. On the contrary, there is no common definition for the end of the boxplots whiskers. In the present work we adopt the following formalism: outliers are defined as those data points lying outside the range ($Q_1 - 1.5 \times IQR$; $Q_3 + 1.5 \times IQR$); extreme events are defined as those data points exceeding the range ($Q_1 - 3 \times IQR$; $Q_3 + 3\times IQR$). 
Several graphical solutions for boxplots are present nowadays, and  generalized versions allow to apply them to skewed distributions, also, by assuring a robust measure of the skewness in the determination of the whiskers\cite{vandervieren2004adjusted}. 
We recall here that the quartiles are also called quantiles of order $1/4$, $1/2$, $3/4$, or $Q_{\frac{1}{4}}$, $Q_{\frac{1}{2}}$, and $Q_{\frac{3}{4}}$, respectively. That second formalism will be used along the paper.

\subsection*{Building the graph: projection in the space of plants/VOCs}
Data are represented as an undirected bipartite graph $G(N,E)$, where every plant species $p$ is connected to its features, i.e. in that case the VOCs it emits. No connection is present between the two set of nodes, i.e. the plant species and the recorded VOCs. Usually, a bipartite graph can also be described by a binary matrix $A(p,f)$ whose element $a_{ij}$ is $1$ just if plant $p$ shows the feature $f$. The most immediate way to measure correlation between species is counting how many VOCs the plants species share in term of significant emissions, and similarly how many plants emit the same VOCs. We refer to the Basic Network Analysis subsection for a proper description of the methodology.
In formulas, this corresponds to consider the matrix of species $P(p,p)=AA^T$ and the matrix of volatile organic compounds, $F(f,f)=A^TA$, i.e. the two bipartite projections of $G(N,E)$.
In the present work, we focused on the graph having as nodes the different plants, i.e. on the {\em Plants graph} $G^P(N,E)$ whose edges weights are proportional to the number of commonly emitted VOCs between plants. Second, in order to catch the predominant similarities in terms of volatile organic compounds emissions, we analysed the second bipartite projection, i.e. the {\em Features graph}, $G^F(N,E)$, whose nodes represent the emitted VOCs. In that case edges weights were proportional to the number of plants sharing the same emitted compound. 

\subsection*{Basic network analysis}
As regards network analysis, we computed some global and local basic metrics described hereafter.
\begin{itemize}
\item \textit{Graph density} ($D$) is defined as the ratio between the numbers of existing edges and the possible number of edges. Given a $N$-order network, graph density is computed as $D = \frac{2E}{N(N-1)}$. Strictly connected to $D$, is the graph average degree $\overline{k}=\frac{1}{V}\sum_{i=1}^{V}k_i=\frac {2E}{V}$, where $k_i$ is the degree of each vertex in $V$, i.e. the number of edges incident to it.
\item \textit{Network clustering coefficient} ($c$) is the overall measure of clustering in a undirected graph in terms of probability that the adjacent vertices of a vertex are connected. More intuitively, global clustering coefficient is simply the ratio of the triangles and the connected triples in the graph. The corresponding local metric is the \textit{local clustering coefficient}, which is the tendency among two vertices to be connected if they share a mutual neighbour. In this analysis we used a local vertex-level quantity\cite{barrat2004architecture} defined in Eq.~\eqref{eq:weighted_C}: 
\begin{equation}\label{eq:weighted_C}
c_{i}^{w} = \frac{1}{s_{i}(k_{i}-1)} \sum_{jh} \frac{(w_{ij}+w_{ih})}{2}a_{ij}a_{ih}a_{jh},
\end{equation}
The normalization factor $\frac{1}{s_{i}(k_{i}-1)}$ accounts for the weight of each edge times the maximum possible
number of triplets in which it may participate, and it ensures that $0 \leq c_{i}^{w} \leq 1$.
That metric combines the topological information with the weight distribution of the network, and it is a measure of the local cohesiveness, grounding on the importance of the clustered structure evaluated on the basis of the amount of interaction intensity actually found on the local triplets\cite{barrat2004architecture}.
\item \textit{Network strength} ($s$) is obtained by summing up the edge weights of the adjacent edges for each vertex\cite{barrat2004architecture}. That metric is a more significant measure of the network properties in terms of the actual weights, and is obtained by extending the definition of \textit{vertex degree} $k_{i} = \sum_{j}a_{ij}$, with $a_{ij}$ elements of the network adjacent matrix $\textbf{A}$. In formulas, $s_{i} = \sum_{j = 1}^{N}a_{ij}w_{if}$.
\end{itemize}

\subsection*{Grouping plants from graph: communities detection analysis} \label{sec:appendix_cd}
Communities detection aims essentially at determine a finite set of categories (clusters or communities) able to describe a data set, according to similarities among its objects\cite{campello2007fuzzy}. More in general, hierarchy is a central organising principle of complex networks, able to offer insight into many complex network phenomena\cite{clauset2008hierarchical}. 
In the present work we adopted the following methods belonging to complex networks framework: 
\begin{itemize}
\item {\bf Fast greedy (FG)} hierarchical agglomeration algorithm\cite{clauset2004finding} is a faster version of the previous greedy optimisation of modularity\cite{newman2004finding}. FG gives identical results in terms of found communities. However, by exploiting some shortcuts in the optimisation problem and using more sophisticated data structures, it runs far more quickly, in time $O(md\log n)$, where $d$ is the depth of the ``dendrogram'' describing the network community structure.
\item {\bf Walktrap community finding algorithm (WT)} finds densely connected subgraphs from a undirected locally dense graph \textit{via} random walks. The basic idea is that short random walks tend to stay in the same community\cite{pons2005computing}. Starting from this point, $WT$ is a measure of similarities between vertices based on random walks, which captures well the community structure in a network, working at various scales. Computation is efficient and the method can be used in an agglomerative algorithm to compute efficiently the community structure of a network.
\item {\bf Louvain or Blondel method (BL)} \cite{blondel2008fast} to uncover modular communities in large networks requiring a coarse-grained description. \textit{Louvain} method ($BL$) is an heuristic approach based on the optimisation of the modularity parameter ($Q$) to infer hierarchical organization. Modularity (Eq.~\eqref{eq:Modularity}) measures the strength of a network division into modules\cite{newman2004finding,newman2004fast}, as it follows:
\begin{equation}
\label{eq:Modularity}
Q = \frac{1}{2m} \sum_{vw} \left[A_{vw} - \frac{k_{v}k_{w}}{ \left(2m \right)} \right]\delta\left(c_{v}, c_{w}\right) = \sum^{c}_{i=1}(e_{ii} - a^{2}_{i}),
\end{equation}
where, $e_{ii}$ is the fraction of edges which connect vertices both lying in the same community $i$, and $a_{i}$ is the fraction of ends of edges that connect vertices in community $i$, in formulas: $e_{ii} = \frac{1}{2m}\sum_{vw} \left[A_{vw} \delta\left(c_{v}, c_{w}\right) \right]$, and $a_{i} = \frac{k_{i}}{2m}=\sum_{i}e_{ij}$; $\textbf{A}$ is the adjacent matrix for the network; $c$ the number of communities; $k_{i} = \sum_{w} A_{vw}$ the degree of the vertex-$i$, $n$ and $m = \frac{1}{2}\sum_{vw} A_{vw}$ the number of graph vertices and edges, respectively. \textit{Delta} function, $\delta(i,j)$, is $1$ if $i=j$, and 0 otherwise.	
\item {\bf Label propagation (LP)} community detection method is a fast, nearly linear time algorithm for detecting community structure in networks\cite{raghavan2007near}. Vertices are initialised with a unique label and, at every step, each node adopts the label that most of its neighbours currently have, that is by a process similar to an `updating by majority voting' in the neighbourhood of the vertex. Moreover, $LP$ uses the network structure alone to run, without requiring neither optimisation of a predefined objective function nor \textit{a-priori} information about the communities, thus overcoming the usual big limitation of having communities which are implicitly defined by the specific algorithm adopted, without an explicit definition. In this iterative process densely connected groups of nodes form a consensus on a unique label to form communities.
\end{itemize}
Besides the complex networks communities detection methodologies, a classic cluster analysis\cite{jolliffe2002principal,macqueen1967some} based on dimensionality reduction methods was also performed to assure the results robustness and reliability, by rejecting those solutions not independent from the statistical methodology applied.


\newpage
\begin{thebibliography}{10}
	\expandafter\ifx\csname url\endcsname\relax
	\def\url#1{\texttt{#1}}\fi
	\expandafter\ifx\csname urlprefix\endcsname\relax\def\urlprefix{URL }\fi
	\providecommand{\bibinfo}[2]{#2}
	\providecommand{\eprint}[2][]{\url{#2}}
	
	\bibitem{theis2003evolution}
	\bibinfo{author}{Theis, N.} \& \bibinfo{author}{Lerdau, M.}
	\newblock \bibinfo{title}{The evolution of function in plant secondary
		metabolites}.
	\newblock \emph{\bibinfo{journal}{International Journal of Plant Sciences}}
	\textbf{\bibinfo{volume}{164}}, \bibinfo{pages}{S93--S102}
	(\bibinfo{year}{2003}).
	
	\bibitem{pichersky2000genetics}
	\bibinfo{author}{Pichersky, E.} \& \bibinfo{author}{Gang, D.~R.}
	\newblock \bibinfo{title}{Genetics and biochemistry of secondary metabolites in
		plants: an evolutionary perspective}.
	\newblock \emph{\bibinfo{journal}{Trends in plant science}}
	\textbf{\bibinfo{volume}{5}}, \bibinfo{pages}{439--445}
	(\bibinfo{year}{2000}).
	
	\bibitem{dicke2010induced}
	\bibinfo{author}{Dicke, M.} \& \bibinfo{author}{Loreto, F.}
	\newblock \bibinfo{title}{Induced plant volatiles: from genes to climate
		change}.
	\newblock \emph{\bibinfo{journal}{Trends in plant science}}
	\textbf{\bibinfo{volume}{15}}, \bibinfo{pages}{115} (\bibinfo{year}{2010}).
	
	\bibitem{dudareva2006plant}
	\bibinfo{author}{Dudareva, N.}, \bibinfo{author}{Negre, F.},
	\bibinfo{author}{Nagegowda, D.~A.} \& \bibinfo{author}{Orlova, I.}
	\newblock \bibinfo{title}{Plant volatiles: recent advances and future
		perspectives}.
	\newblock \emph{\bibinfo{journal}{Critical reviews in plant sciences}}
	\textbf{\bibinfo{volume}{25}}, \bibinfo{pages}{417--440}
	(\bibinfo{year}{2006}).
	
	\bibitem{penuelas2001complexity}
	\bibinfo{author}{Pe{\~n}uelas, J.} \& \bibinfo{author}{Llusia, J.}
	\newblock \bibinfo{title}{The complexity of factors driving volatile organic
		compound emissions by plants}.
	\newblock \emph{\bibinfo{journal}{Biologia Plantarum}}
	\textbf{\bibinfo{volume}{44}}, \bibinfo{pages}{481--487}
	(\bibinfo{year}{2001}).
	
	\bibitem{holopainen2010leaf}
	\bibinfo{author}{Holopainen, J.~K.}, \bibinfo{author}{Heijari, J.},
	\bibinfo{author}{Oksanen, E.} \& \bibinfo{author}{Alessio, G.~A.}
	\newblock \bibinfo{title}{Leaf volatile emissions of betula pendula during
		autumn coloration and leaf fall}.
	\newblock \emph{\bibinfo{journal}{Journal of chemical ecology}}
	\textbf{\bibinfo{volume}{36}}, \bibinfo{pages}{1068--1075}
	(\bibinfo{year}{2010}).
	
	\bibitem{holopainen2010multiple}
	\bibinfo{author}{Holopainen, J.~K.} \& \bibinfo{author}{Gershenzon, J.}
	\newblock \bibinfo{title}{Multiple stress factors and the emission of plant
		vocs}.
	\newblock \emph{\bibinfo{journal}{Trends in plant science}}
	\textbf{\bibinfo{volume}{15}}, \bibinfo{pages}{176--184}
	(\bibinfo{year}{2010}).
	
	\bibitem{spinelli2011emission}
	\bibinfo{author}{Spinelli, F.}, \bibinfo{author}{Cellini, A.},
	\bibinfo{author}{Piovene, C.}, \bibinfo{author}{Nagesh, K.~M.} \&
	\bibinfo{author}{Marchetti, L.}
	\newblock \emph{\bibinfo{title}{Emission and function of volatile organic
			compounds in response to abiotic stress}} (\bibinfo{publisher}{INTECH Open
		Access Publisher}, \bibinfo{year}{2011}).
	
	\bibitem{mumm2003chemical}
	\bibinfo{author}{Mumm, R.}, \bibinfo{author}{Schrank, K.},
	\bibinfo{author}{Wegener, R.}, \bibinfo{author}{Schulz, S.} \&
	\bibinfo{author}{Hilker, M.}
	\newblock \bibinfo{title}{Chemical analysis of volatiles emitted by pinus
		sylvestris after induction by insect oviposition}.
	\newblock \emph{\bibinfo{journal}{Journal of chemical ecology}}
	\textbf{\bibinfo{volume}{29}}, \bibinfo{pages}{1235--1252}
	(\bibinfo{year}{2003}).
	
	\bibitem{dudareva2000biochemical}
	\bibinfo{author}{Dudareva, N.} \& \bibinfo{author}{Pichersky, E.}
	\newblock \bibinfo{title}{Biochemical and molecular genetic aspects of floral
		scents}.
	\newblock \emph{\bibinfo{journal}{Plant physiology}}
	\textbf{\bibinfo{volume}{122}}, \bibinfo{pages}{627--634}
	(\bibinfo{year}{2000}).
	
	\bibitem{baldwin2006volatile}
	\bibinfo{author}{Baldwin, I.~T.}, \bibinfo{author}{Halitschke, R.},
	\bibinfo{author}{Paschold, A.}, \bibinfo{author}{Von~Dahl, C.~C.} \&
	\bibinfo{author}{Preston, C.~A.}
	\newblock \bibinfo{title}{Volatile signaling in plant-plant interactions:"
		talking trees" in the genomics era}.
	\newblock \emph{\bibinfo{journal}{Science}} \textbf{\bibinfo{volume}{311}},
	\bibinfo{pages}{812--815} (\bibinfo{year}{2006}).
	
	\bibitem{heil2010explaining}
	\bibinfo{author}{Heil, M.} \& \bibinfo{author}{Karban, R.}
	\newblock \bibinfo{title}{Explaining evolution of plant communication by
		airborne signals}.
	\newblock \emph{\bibinfo{journal}{Trends in ecology \& evolution}}
	\textbf{\bibinfo{volume}{25}}, \bibinfo{pages}{137--144}
	(\bibinfo{year}{2010}).
	
	\bibitem{war2012mechanisms}
	\bibinfo{author}{War, A.~R.} \emph{et~al.}
	\newblock \bibinfo{title}{Mechanisms of plant defense against insect
		herbivores}.
	\newblock \emph{\bibinfo{journal}{Plant signaling \& behavior}}
	\textbf{\bibinfo{volume}{7}}, \bibinfo{pages}{1306--1320}
	(\bibinfo{year}{2012}).
	
	\bibitem{ruuskanen2009measurements}
	\bibinfo{author}{Ruuskanen, T.} \emph{et~al.}
	\newblock \emph{\bibinfo{title}{Measurements of Volatile Organic Compounds-from
			Biogenic Emissions to Concentrations in Ambient Air}}.
	\newblock Ph.D. thesis, \bibinfo{school}{University of Helsinki, Faculty of
		Science, Department of Physics, Division of Atmospheric Sciences and
		Geophysics} (\bibinfo{year}{2009}).
	
	\bibitem{agrawal2011current}
	\bibinfo{author}{Agrawal, A.~A.}
	\newblock \bibinfo{title}{Current trends in the evolutionary ecology of plant
		defence}.
	\newblock \emph{\bibinfo{journal}{Functional Ecology}}
	\textbf{\bibinfo{volume}{25}}, \bibinfo{pages}{420--432}
	(\bibinfo{year}{2011}).
	
	\bibitem{berenbaum2008facing}
	\bibinfo{author}{Berenbaum, M.~R.} \& \bibinfo{author}{Zangerl, A.~R.}
	\newblock \bibinfo{title}{Facing the future of plant-insect interaction
		research: le retour {\`a} la ''raison d'{\^e}tre''}.
	\newblock \emph{\bibinfo{journal}{Plant Physiology}}
	\textbf{\bibinfo{volume}{146}}, \bibinfo{pages}{804--811}
	(\bibinfo{year}{2008}).
	
	\bibitem{llusia2002seasonal}
	\bibinfo{author}{Llusia, J.}, \bibinfo{author}{Penuelas, J.} \&
	\bibinfo{author}{Gimeno, B.}
	\newblock \bibinfo{title}{Seasonal and species-specific response of voc
		emissions by mediterranean woody plant to elevated ozone concentrations}.
	\newblock \emph{\bibinfo{journal}{Atmospheric Environment}}
	\textbf{\bibinfo{volume}{36}}, \bibinfo{pages}{3931--3938}
	(\bibinfo{year}{2002}).
	
	\bibitem{caldarelli2007scale}
	\bibinfo{author}{Caldarelli, G.}
	\newblock \bibinfo{title}{{Scale-Free Networks: complex webs in nature and
			technology}}.
	\newblock \emph{\bibinfo{journal}{OUP Catalogue}}  (\bibinfo{year}{2007}).
	
	\bibitem{Boccaletti2006175}
	\bibinfo{author}{Boccaletti, S.}, \bibinfo{author}{Latora, V.},
	\bibinfo{author}{Moreno, Y.}, \bibinfo{author}{Chavez, M.} \&
	\bibinfo{author}{Hwang, D.~U.}
	\newblock \bibinfo{title}{Complex networks: Structure and dynamics}.
	\newblock \emph{\bibinfo{journal}{Physics Reports}}
	\textbf{\bibinfo{volume}{424}}, \bibinfo{pages}{175 -- 308}
	(\bibinfo{year}{2006}).
	\newblock
	\urlprefix\url{http://www.sciencedirect.com/science/article/pii/S037015730500462X}.
	
	\bibitem{barrat2004architecture}
	\bibinfo{author}{Barrat, A.}, \bibinfo{author}{Barthelemy, M.},
	\bibinfo{author}{Pastor-Satorras, R.} \& \bibinfo{author}{Vespignani, A.}
	\newblock \bibinfo{title}{The architecture of complex weighted networks}.
	\newblock \emph{\bibinfo{journal}{Proceedings of the National Academy of
			Sciences of the United States of America}} \textbf{\bibinfo{volume}{101}},
	\bibinfo{pages}{3747--3752} (\bibinfo{year}{2004}).
	
	\bibitem{raghavan2007near}
	\bibinfo{author}{Raghavan, U.~N.}, \bibinfo{author}{Albert, R.} \&
	\bibinfo{author}{Kumara, S.}
	\newblock \bibinfo{title}{Near linear time algorithm to detect community
		structures in large-scale networks}.
	\newblock \emph{\bibinfo{journal}{Physical Review E}}
	\textbf{\bibinfo{volume}{76}}, \bibinfo{pages}{036106}
	(\bibinfo{year}{2007}).
	
	\bibitem{newman2004finding}
	\bibinfo{author}{Newman, M.~E.} \& \bibinfo{author}{Girvan, M.}
	\newblock \bibinfo{title}{Finding and evaluating community structure in
		networks}.
	\newblock \emph{\bibinfo{journal}{Physical review E}}
	\textbf{\bibinfo{volume}{69}}, \bibinfo{pages}{026113}
	(\bibinfo{year}{2004}).
	
	\bibitem{ma2016wiener}
	\bibinfo{author}{Ma, J.}, \bibinfo{author}{Shi, Y.}, \bibinfo{author}{Wang, Z.}
	\& \bibinfo{author}{Yue, J.}
	\newblock \bibinfo{title}{On wiener polarity index of bicyclic networks}.
	\newblock \emph{\bibinfo{journal}{Scientific reports}}
	\textbf{\bibinfo{volume}{6}} (\bibinfo{year}{2016}).
	
	\bibitem{li2013note}
	\bibinfo{author}{Li, X.}, \bibinfo{author}{Li, Y.}, \bibinfo{author}{Shi, Y.}
	\& \bibinfo{author}{Gutman, I.}
	\newblock \bibinfo{title}{Note on the homo-lumo index of graphs}.
	\newblock \emph{\bibinfo{journal}{MATCH Commun. Math. Comput. Chem}}
	\textbf{\bibinfo{volume}{70}}, \bibinfo{pages}{85--96}
	(\bibinfo{year}{2013}).
	
	\bibitem{cao2014extremality}
	\bibinfo{author}{Cao, S.}, \bibinfo{author}{Dehmer, M.} \&
	\bibinfo{author}{Shi, Y.}
	\newblock \bibinfo{title}{Extremality of degree-based graph entropies}.
	\newblock \emph{\bibinfo{journal}{Information Sciences}}
	\textbf{\bibinfo{volume}{278}}, \bibinfo{pages}{22--33}
	(\bibinfo{year}{2014}).
	
	\bibitem{Dunne01102002}
	\bibinfo{author}{Dunne, J.~A.}, \bibinfo{author}{Williams, R.~J.} \&
	\bibinfo{author}{Martinez, N.~D.}
	\newblock \bibinfo{title}{Food-web structure and network theory: The role of
		connectance and size}.
	\newblock \emph{\bibinfo{journal}{Proceedings of the National Academy of
			Sciences}} \textbf{\bibinfo{volume}{99}}, \bibinfo{pages}{12917--12922}
	(\bibinfo{year}{2002}).
	\newblock \urlprefix\url{http://www.pnas.org/content/99/20/12917.abstract}.
	\newblock \eprint{http://www.pnas.org/content/99/20/12917.full.pdf}.
	
	\bibitem{Stelzl2005957}
	\bibinfo{author}{Stelzl, U.} \emph{et~al.}
	\newblock \bibinfo{title}{A human protein-protein interaction network: A
		resource for annotating the proteome}.
	\newblock \emph{\bibinfo{journal}{Cell}} \textbf{\bibinfo{volume}{122}},
	\bibinfo{pages}{957 -- 968} (\bibinfo{year}{2005}).
	\newblock
	\urlprefix\url{http://www.sciencedirect.com/science/article/pii/S0092867405008664}.
	
	\bibitem{proulx2005network}
	\bibinfo{author}{Proulx, S.~R.}, \bibinfo{author}{Promislow, D.~E.} \&
	\bibinfo{author}{Phillips, P.~C.}
	\newblock \bibinfo{title}{Network thinking in ecology and evolution}.
	\newblock \emph{\bibinfo{journal}{Trends in Ecology \& Evolution}}
	\textbf{\bibinfo{volume}{20}}, \bibinfo{pages}{345--353}
	(\bibinfo{year}{2005}).
	
	\bibitem{barabasi2011network}
	\bibinfo{author}{Barab{\'a}si, A.-L.}, \bibinfo{author}{Gulbahce, N.} \&
	\bibinfo{author}{Loscalzo, J.}
	\newblock \bibinfo{title}{Network medicine: a network-based approach to human
		disease}.
	\newblock \emph{\bibinfo{journal}{Nature Reviews Genetics}}
	\textbf{\bibinfo{volume}{12}}, \bibinfo{pages}{56--68}
	(\bibinfo{year}{2011}).
	
	\bibitem{leecomorbidity}
	\bibinfo{author}{Lee, D.-S.} \emph{et~al.}
	\newblock \bibinfo{title}{The implications of human metabolic network topology
		for disease comorbidity}.
	\newblock \emph{\bibinfo{journal}{Proceedings of the National Academy of
			Sciences of the United States of America}} \textbf{\bibinfo{volume}{105}},
	\bibinfo{pages}{9880--9885} (\bibinfo{year}{2008}).
	
	\bibitem{stephan2000computational}
	\bibinfo{author}{Stephan, K.~E.} \emph{et~al.}
	\newblock \bibinfo{title}{Computational analysis of functional connectivity
		between areas of primate cerebral cortex}.
	\newblock \emph{\bibinfo{journal}{Philosophical Transactions of the Royal
			Society of London B: Biological Sciences}} \textbf{\bibinfo{volume}{355}},
	\bibinfo{pages}{111--126} (\bibinfo{year}{2000}).
	
	\bibitem{caretta2008}
	\bibinfo{author}{Caretta~Cartozo, C.}, \bibinfo{author}{Garlaschelli, D.},
	\bibinfo{author}{Ricotta, C.}, \bibinfo{author}{M., B.} \&
	\bibinfo{author}{G., C.}
	\newblock \bibinfo{title}{Quantifying the universal taxonomic diversity in real
		species assemblage}.
	\newblock \emph{\bibinfo{journal}{Journal of Physics A}}
	\textbf{\bibinfo{volume}{41}}, \bibinfo{pages}{224012}
	(\bibinfo{year}{2008}).
	
	\bibitem{vivaldo2016networks}
	\bibinfo{author}{Vivaldo, G.}, \bibinfo{author}{Masi, E.},
	\bibinfo{author}{Pandolfi, C.}, \bibinfo{author}{Mancuso, S.} \&
	\bibinfo{author}{Caldarelli, G.}
	\newblock \bibinfo{title}{Networks of plants: how to measure similarity in
		vegetable species}.
	\newblock \emph{\bibinfo{journal}{arXiv preprint arXiv:1602.05887}}
	(\bibinfo{year}{2016}).
	
	\bibitem{tukey1977exploratory}
	\bibinfo{author}{Tukey, J.}
	\newblock \bibinfo{title}{Exploratory data analysis.-reading, mass.:
		Addison-wesley}.
	\newblock \emph{\bibinfo{journal}{Exploratory data analysis: Reading, Mass:
			Addison-Wesley}}  (\bibinfo{year}{1977}).
	
	\bibitem{vandervieren2004adjusted}
	\bibinfo{author}{Vandervieren, E.} \& \bibinfo{author}{Hubert, M.}
	\newblock \bibinfo{title}{An adjusted boxplot for skewed distributions}.
	\newblock \emph{\bibinfo{journal}{COMPSTAT 2004, proceedings in computational
			statistics. Springer, Heidelberg}} \bibinfo{pages}{1933--1940}
	(\bibinfo{year}{2004}).
	
	\bibitem{loreto2006induction}
	\bibinfo{author}{Loreto, F.}, \bibinfo{author}{Barta, C.},
	\bibinfo{author}{Brilli, F.} \& \bibinfo{author}{Nogues, I.}
	\newblock \bibinfo{title}{On the induction of volatile organic compound
		emissions by plants as consequence of wounding or fluctuations of light and
		temperature}.
	\newblock \emph{\bibinfo{journal}{Plant, cell \& environment}}
	\textbf{\bibinfo{volume}{29}}, \bibinfo{pages}{1820--1828}
	(\bibinfo{year}{2006}).
	
	\bibitem{brilli2011detection}
	\bibinfo{author}{Brilli, F.} \emph{et~al.}
	\newblock \bibinfo{title}{Detection of plant volatiles after leaf wounding and
		darkening by proton transfer reaction 'time-of-flight' mass spectrometry
		(ptr-tof)}.
	\newblock \emph{\bibinfo{journal}{PLoS One}} \textbf{\bibinfo{volume}{6}},
	\bibinfo{pages}{e20419} (\bibinfo{year}{2011}).
	
	\bibitem{degen2004high}
	\bibinfo{author}{Degen, T.}, \bibinfo{author}{Dillmann, C.},
	\bibinfo{author}{Marion-Poll, F.} \& \bibinfo{author}{Turlings, T.~C.}
	\newblock \bibinfo{title}{High genetic variability of herbivore-induced
		volatile emission within a broad range of maize inbred lines}.
	\newblock \emph{\bibinfo{journal}{Plant physiology}}
	\textbf{\bibinfo{volume}{135}}, \bibinfo{pages}{1928--1938}
	(\bibinfo{year}{2004}).
	
	\bibitem{wu2008comparison}
	\bibinfo{author}{Wu, J.}, \bibinfo{author}{Hettenhausen, C.},
	\bibinfo{author}{Schuman, M.~C.} \& \bibinfo{author}{Baldwin, I.~T.}
	\newblock \bibinfo{title}{A comparison of two nicotiana attenuata accessions
		reveals large differences in signaling induced by oral secretions of the
		specialist herbivore manduca sexta}.
	\newblock \emph{\bibinfo{journal}{Plant Physiology}}
	\textbf{\bibinfo{volume}{146}}, \bibinfo{pages}{927--939}
	(\bibinfo{year}{2008}).
	
	\bibitem{van1991identification}
	\bibinfo{author}{Van~Langenhove, H.~J.}, \bibinfo{author}{Cornelis, C.~P.} \&
	\bibinfo{author}{Schamp, N.~M.}
	\newblock \bibinfo{title}{Identification of volatiles emitted during the
		blanching process of brussels sprouts and cauliflower}.
	\newblock \emph{\bibinfo{journal}{Journal of the Science of Food and
			Agriculture}} \textbf{\bibinfo{volume}{55}}, \bibinfo{pages}{483--487}
	(\bibinfo{year}{1991}).
	
	\bibitem{geervliet1997comparative}
	\bibinfo{author}{Geervliet, J.~B.}, \bibinfo{author}{Posthumus, M.~A.},
	\bibinfo{author}{Vet, L.~E.} \& \bibinfo{author}{Dicke, M.}
	\newblock \bibinfo{title}{Comparative analysis of headspace volatiles from
		different caterpillar-infested or uninfested food plants of pieris species}.
	\newblock \emph{\bibinfo{journal}{Journal of chemical ecology}}
	\textbf{\bibinfo{volume}{23}}, \bibinfo{pages}{2935--2954}
	(\bibinfo{year}{1997}).
	
	\bibitem{buhr2002analysis}
	\bibinfo{author}{Buhr, K.}, \bibinfo{author}{van Ruth, S.} \&
	\bibinfo{author}{Delahunty, C.}
	\newblock \bibinfo{title}{Analysis of volatile flavour compounds by proton
		transfer reaction-mass spectrometry: fragmentation patterns and
		discrimination between isobaric and isomeric compounds}.
	\newblock \emph{\bibinfo{journal}{International Journal of Mass Spectrometry}}
	\textbf{\bibinfo{volume}{221}}, \bibinfo{pages}{1--7} (\bibinfo{year}{2002}).
	
	\bibitem{pierre2011differences}
	\bibinfo{author}{Pierre, P.~S.} \emph{et~al.}
	\newblock \bibinfo{title}{Differences in volatile profiles of turnip plants
		subjected to single and dual herbivory above-and belowground}.
	\newblock \emph{\bibinfo{journal}{Journal of chemical ecology}}
	\textbf{\bibinfo{volume}{37}}, \bibinfo{pages}{368--377}
	(\bibinfo{year}{2011}).
	
	\bibitem{dudareva2013biosynthesis}
	\bibinfo{author}{Dudareva, N.}, \bibinfo{author}{Klempien, A.},
	\bibinfo{author}{Muhlemann, J.~K.} \& \bibinfo{author}{Kaplan, I.}
	\newblock \bibinfo{title}{Biosynthesis, function and metabolic engineering of
		plant volatile organic compounds}.
	\newblock \emph{\bibinfo{journal}{New Phytologist}}
	\textbf{\bibinfo{volume}{198}}, \bibinfo{pages}{16--32}
	(\bibinfo{year}{2013}).
	
	\bibitem{sun2016my}
	\bibinfo{author}{Sun, P.}, \bibinfo{author}{Schuurink, R.~C.},
	\bibinfo{author}{Caissard, J.-C.}, \bibinfo{author}{Hugueney, P.} \&
	\bibinfo{author}{Baudino, S.}
	\newblock \bibinfo{title}{My way: Noncanonical biosynthesis pathways for plant
		volatiles}.
	\newblock \emph{\bibinfo{journal}{Trends in Plant Science}}
	(\bibinfo{year}{2016}).
	
	\bibitem{lindfors2000biogenic}
	\bibinfo{author}{Lindfors, V.} \& \bibinfo{author}{Laurila, T.}
	\newblock \bibinfo{title}{Biogenic volatile organic compound (voc) emissions
		from forests in finland}.
	\newblock \emph{\bibinfo{journal}{Boreal environment research}}
	\textbf{\bibinfo{volume}{5}}, \bibinfo{pages}{95--113}
	(\bibinfo{year}{2000}).
	
	\bibitem{bentley2004environmental}
	\bibinfo{author}{Bentley, R.} \& \bibinfo{author}{Chasteen, T.~G.}
	\newblock \bibinfo{title}{Environmental voscs----formation and degradation of
		dimethyl sulfide, methanethiol and related materials}.
	\newblock \emph{\bibinfo{journal}{Chemosphere}} \textbf{\bibinfo{volume}{55}},
	\bibinfo{pages}{291--317} (\bibinfo{year}{2004}).
	
	\bibitem{dobson2006relationship}
	\bibinfo{author}{Dobson, H.~E.}
	\newblock \bibinfo{title}{Relationship between floral fragrance composition and
		type of pollinator}.
	\newblock \emph{\bibinfo{journal}{Biology of floral scent}}
	\bibinfo{pages}{147--198} (\bibinfo{year}{2006}).
	
	\bibitem{kumari2014essoildb}
	\bibinfo{author}{Kumari, S.} \emph{et~al.}
	\newblock \bibinfo{title}{Essoildb: a database of essential oils reflecting
		terpene composition and variability in the plant kingdom}.
	\newblock \emph{\bibinfo{journal}{Database}} \textbf{\bibinfo{volume}{2014}},
	\bibinfo{pages}{bau120} (\bibinfo{year}{2014}).
	
	\bibitem{edris2007pharmaceutical}
	\bibinfo{author}{Edris, A.~E.}
	\newblock \bibinfo{title}{Pharmaceutical and therapeutic potentials of
		essential oils and their individual volatile constituents: a review}.
	\newblock \emph{\bibinfo{journal}{Phytotherapy research}}
	\textbf{\bibinfo{volume}{21}}, \bibinfo{pages}{308--323}
	(\bibinfo{year}{2007}).
	
	\bibitem{lindinger1998proton}
	\bibinfo{author}{Lindinger, W.} \& \bibinfo{author}{Jordan, A.}
	\newblock \bibinfo{title}{Proton-transfer-reaction mass spectrometry (ptr--ms):
		on-line monitoring of volatile organic compounds at pptv levels}.
	\newblock \emph{\bibinfo{journal}{Chemical Society Reviews}}
	\textbf{\bibinfo{volume}{27}}, \bibinfo{pages}{347--375}
	(\bibinfo{year}{1998}).
	
	\bibitem{jordan2009high}
	\bibinfo{author}{Jordan, A.} \emph{et~al.}
	\newblock \bibinfo{title}{A high resolution and high sensitivity
		proton-transfer-reaction time-of-flight mass spectrometer (ptr-tof-ms)}.
	\newblock \emph{\bibinfo{journal}{International Journal of Mass Spectrometry}}
	\textbf{\bibinfo{volume}{286}}, \bibinfo{pages}{122--128}
	(\bibinfo{year}{2009}).
	
	\bibitem{taiti2016assessing}
	\bibinfo{author}{Taiti, C.} \emph{et~al.}
	\newblock \bibinfo{title}{Assessing voc emission by different wood cores using
		the ptr-tof-ms technology}.
	\newblock \emph{\bibinfo{journal}{Wood Science and Technology}}
	\bibinfo{pages}{1--23}.
	
	\bibitem{mancuso2015soil}
	\bibinfo{author}{Mancuso, S.} \emph{et~al.}
	\newblock \bibinfo{title}{Soil volatile analysis by proton transfer
		reaction-time of flight mass spectrometry (ptr-tof-ms)}.
	\newblock \emph{\bibinfo{journal}{Applied Soil Ecology}}
	\textbf{\bibinfo{volume}{86}}, \bibinfo{pages}{182--191}
	(\bibinfo{year}{2015}).
	
	\bibitem{pang2015biogenic}
	\bibinfo{author}{Pang, X.}
	\newblock \bibinfo{title}{Biogenic volatile organic compound analyses by
		ptr-tof-ms: Calibration, humidity effect and reduced electric field
		dependency}.
	\newblock \emph{\bibinfo{journal}{Journal of Environmental Sciences}}
	\textbf{\bibinfo{volume}{32}}, \bibinfo{pages}{196--206}
	(\bibinfo{year}{2015}).
	
	\bibitem{herbig2009line}
	\bibinfo{author}{Herbig, J.} \emph{et~al.}
	\newblock \bibinfo{title}{On-line breath analysis with ptr-tof}.
	\newblock \emph{\bibinfo{journal}{Journal of breath research}}
	\textbf{\bibinfo{volume}{3}}, \bibinfo{pages}{027004} (\bibinfo{year}{2009}).
	
	\bibitem{taiti2016sometimes}
	\bibinfo{author}{Taiti, C.} \emph{et~al.}
	\newblock \bibinfo{title}{Sometimes a little mango goes a long way: A rapid
		approach to assess how different shipping systems affect fruit commercial
		quality}.
	\newblock \emph{\bibinfo{journal}{Food analytical methods}}
	\textbf{\bibinfo{volume}{9}}, \bibinfo{pages}{691--698}
	(\bibinfo{year}{2016}).
	
	\bibitem{lanza2015selective}
	\bibinfo{author}{Lanza, M.} \emph{et~al.}
	\newblock \bibinfo{title}{Selective reagent ionisation-time of flight-mass
		spectrometry: a rapid technology for the novel analysis of blends of new
		psychoactive substances}.
	\newblock \emph{\bibinfo{journal}{Journal of Mass Spectrometry}}
	\textbf{\bibinfo{volume}{50}}, \bibinfo{pages}{427--431}
	(\bibinfo{year}{2015}).
	
	\bibitem{maleknia2007ptr}
	\bibinfo{author}{Maleknia, S.~D.}, \bibinfo{author}{Bell, T.~L.} \&
	\bibinfo{author}{Adams, M.~A.}
	\newblock \bibinfo{title}{Ptr-ms analysis of reference and plant-emitted
		volatile organic compounds}.
	\newblock \emph{\bibinfo{journal}{International Journal of Mass Spectrometry}}
	\textbf{\bibinfo{volume}{262}}, \bibinfo{pages}{203--210}
	(\bibinfo{year}{2007}).
	
	\bibitem{kim2009measurement}
	\bibinfo{author}{Kim, S.} \emph{et~al.}
	\newblock \bibinfo{title}{Measurement of atmospheric sesquiterpenes by proton
		transfer reaction-mass spectrometry (ptr-ms)}.
	\newblock \emph{\bibinfo{journal}{Atmospheric Measurement Techniques}}
	\textbf{\bibinfo{volume}{2}} (\bibinfo{year}{2009}).
	
	\bibitem{demarcke2009laboratory}
	\bibinfo{author}{Demarcke, M.} \emph{et~al.}
	\newblock \bibinfo{title}{Laboratory studies in support of the detection of
		sesquiterpenes by proton-transfer-reaction-mass-spectrometry}.
	\newblock \emph{\bibinfo{journal}{International Journal of Mass Spectrometry}}
	\textbf{\bibinfo{volume}{279}}, \bibinfo{pages}{156--162}
	(\bibinfo{year}{2009}).
	
	\bibitem{papurello2012monitoring}
	\bibinfo{author}{Papurello, D.} \emph{et~al.}
	\newblock \bibinfo{title}{Monitoring of volatile compound emissions during dry
		anaerobic digestion of the organic fraction of municipal solid waste by
		proton transfer reaction time-of-flight mass spectrometry}.
	\newblock \emph{\bibinfo{journal}{Bioresource technology}}
	\textbf{\bibinfo{volume}{126}}, \bibinfo{pages}{254--265}
	(\bibinfo{year}{2012}).
	
	\bibitem{liu2013experimental}
	\bibinfo{author}{Liu, D.}, \bibinfo{author}{Andreasen, R.~R.},
	\bibinfo{author}{Poulsen, T.~G.} \& \bibinfo{author}{Feilberg, A.}
	\newblock \bibinfo{title}{Experimental determination of mass transfer
		coefficients of volatile sulfur odorants in biofilter media measured by
		proton-transfer-reaction mass spectrometry (ptr-ms)}.
	\newblock \emph{\bibinfo{journal}{Chemical engineering journal}}
	\textbf{\bibinfo{volume}{219}}, \bibinfo{pages}{335--345}
	(\bibinfo{year}{2013}).
	
	\bibitem{schwarz2009determining}
	\bibinfo{author}{Schwarz, K.}, \bibinfo{author}{Filipiak, W.} \&
	\bibinfo{author}{Amann, A.}
	\newblock \bibinfo{title}{Determining concentration patterns of volatile
		compounds in exhaled breath by ptr-ms}.
	\newblock \emph{\bibinfo{journal}{Journal of Breath Research}}
	\textbf{\bibinfo{volume}{3}}, \bibinfo{pages}{027002} (\bibinfo{year}{2009}).
	
	\bibitem{soukoulis2013ptr}
	\bibinfo{author}{Soukoulis, C.} \emph{et~al.}
	\newblock \bibinfo{title}{Ptr-tof-ms, a novel, rapid, high sensitivity and
		non-invasive tool to monitor volatile compound release during fruit
		post-harvest storage: the case study of apple ripening}.
	\newblock \emph{\bibinfo{journal}{Food and Bioprocess Technology}}
	\textbf{\bibinfo{volume}{6}}, \bibinfo{pages}{2831--2843}
	(\bibinfo{year}{2013}).
	
	\bibitem{campello2007fuzzy}
	\bibinfo{author}{Campello, R.}
	\newblock \bibinfo{title}{{A Fuzzy Extension of the Rand Index and Other
			Related Indexes for Clustering and Classification Assessment}}.
	\newblock \emph{\bibinfo{journal}{Pattern Recognition Letters,}}
	\textbf{\bibinfo{volume}{28}} (\bibinfo{year}{2007}).
	
	\bibitem{clauset2008hierarchical}
	\bibinfo{author}{Clauset, A.}, \bibinfo{author}{Moore, C.} \&
	\bibinfo{author}{Newman, M.}
	\newblock \bibinfo{title}{{Hierarchical structure and the prediction of missing
			links in networks.}}
	\newblock \emph{\bibinfo{journal}{Nature}} \textbf{\bibinfo{volume}{453}},
	\bibinfo{pages}{98--101} (\bibinfo{year}{2008}).
	\newblock \urlprefix\url{http://dx.doi.org/10.1038/nature06830}.
	
	\bibitem{clauset2004finding}
	\bibinfo{author}{Clauset, A.}, \bibinfo{author}{Newman, M.~E.} \&
	\bibinfo{author}{Moore, C.}
	\newblock \bibinfo{title}{Finding community structure in very large networks}.
	\newblock \emph{\bibinfo{journal}{Physical review E}}
	\textbf{\bibinfo{volume}{70}}, \bibinfo{pages}{066111}
	(\bibinfo{year}{2004}).
	
	\bibitem{pons2005computing}
	\bibinfo{author}{Pons, P.} \& \bibinfo{author}{Latapy, M.}
	\newblock \bibinfo{title}{Computing communities in large networks using random
		walks}.
	\newblock In \emph{\bibinfo{booktitle}{Computer and Information Sciences-ISCIS
			2005}}, \bibinfo{pages}{284--293} (\bibinfo{publisher}{Springer},
	\bibinfo{year}{2005}).
	
	\bibitem{blondel2008fast}
	\bibinfo{author}{Blondel, V.~D.}, \bibinfo{author}{Guillaume, J.-L.},
	\bibinfo{author}{Lambiotte, R.} \& \bibinfo{author}{Lefebvre, E.}
	\newblock \bibinfo{title}{{Fast unfolding of communities in large networks}}.
	\newblock \emph{\bibinfo{journal}{Journal of Statistical Mechanics: Theory and
			Experiment}} \textbf{\bibinfo{volume}{2008}}, \bibinfo{pages}{P10008}
	(\bibinfo{year}{2008}).
	\newblock \urlprefix\url{http://stacks.iop.org/1742-5468/2008/i=10/a=P10008}.
	
	\bibitem{newman2004fast}
	\bibinfo{author}{Newman, M.~E.}
	\newblock \bibinfo{title}{Fast algorithm for detecting community structure in
		networks}.
	\newblock \emph{\bibinfo{journal}{Physical review E}}
	\textbf{\bibinfo{volume}{69}}, \bibinfo{pages}{066133}
	(\bibinfo{year}{2004}).
	
	\bibitem{jolliffe2002principal}
	\bibinfo{author}{Jolliffe, I.}
	\newblock \emph{\bibinfo{title}{Principal component analysis}}
	(\bibinfo{publisher}{Wiley Online Library}, \bibinfo{year}{2002}).
	
	\bibitem{macqueen1967some}
	\bibinfo{author}{MacQueen, J.} \emph{et~al.}
	\newblock \bibinfo{title}{Some methods for classification and analysis of
		multivariate observations}.
	\newblock In \emph{\bibinfo{booktitle}{Proceedings of the fifth Berkeley
			symposium on mathematical statistics and probability}},
	vol.~\bibinfo{volume}{1}, \bibinfo{pages}{281--297}
	(\bibinfo{organization}{Oakland, CA, USA.}, \bibinfo{year}{1967}).
	
\end{thebibliography}

\section*{Acknowledgments}
The authors acknowledge support from EU FET Open Project PLEASED nr. 296582. 
GV and GC also acknowledge EU FET Integrated Project MULTIPLEX nr. 317532. 

\section*{Author contributions statement}
GV, EM, CT, GC and SM contributed equally to the analysis of the dataset and to the interpretation of the results of this analysis, both from
the point of view of Network Theory as well as in terms of biological implications. They also contributed equally to the
writing and reviewing of the manuscript.

\section*{Additional information}
\textbf{Competing financial interests.} The authors declare no competing financial interests.

\end{document}